\documentclass[12pt]{iopart}
\usepackage{graphicx,color}
\usepackage{comment}
\usepackage{amssymb}
\usepackage{lscape}

\begin{document}

\newcommand {\nc} {\newcommand}
\nc {\IR} [1]{\textcolor{red}{#1}}
\nc {\IB} [1]{\textcolor{blue}{#1}}
\nc {\IG} [1]{\textcolor{green}{#1}}

\title{Recent advances in the quantification of uncertainties in reaction theory}

\author{A.E. Lovell}
\address{Nuclear Theory Group, Los Alamos National Laboratory, Los Alamos, NM 87545, USA}
\ead{lovell@lanl.gov}

\author{F.M. Nunes and  M. Catacora-Rios} 

\address{National Superconducting Cyclotron Laboratory and Department of Physics and Astronomy, Michigan State University, East Lansing, MI 48824, USA \\
}
\ead{nunes@nscl.msu.edu}

\author{G.B. King}
\address{Department of Physics, Washington University, St. Louis, MO 63130, USA}

\vspace{10pt}
\begin{indented}
\item[] \today
\end{indented}

\begin{abstract}
Uncertainty quantification has become increasingly more prominent in nuclear physics over the past several years.  In few-body reaction theory, there are four main sources that contribute to the uncertainties in the calculated observables:  the effective potentials, approximations made to the few-body problem, structure functions, and degrees of freedom left out of the model space.  In this work, we illustrate some of the features that can be obtained when modern statistical tools are applied in the context of nuclear reactions. This work consists of a summary of the progress that has been made in quantifying theoretical uncertainties in this domain, focusing primarily on those uncertainties coming from the effective optical potential as well as their propagation within various reaction theories.  We use, as the central example, reactions on the doubly-magic stable nucleus $^{40}$Ca, namely neutron and proton elastic scattering and single-nucleon transfer $^{40}$Ca(d,p)$^{41}$Ca.  First, we show different optimization schemes used to constrain the optical potential from differential cross sections and other experimental constraints; we then discuss how these uncertainties propagate to the transfer cross section, comparing two reaction theories.  %\IR{Afterwards, we estimate uncertainties due to the structure functions when constraining the asymptotic normalization coefficient. Finally, we estimate the additional uncertainty arising from the degrees of freedom left out of the model space using constraints from inelastic scattering. } 
We finish by laying out our future plans.
\end{abstract}

\vspace{2pc}

\section{Motivation}
\label{sec:mot}
%FN (and AL)

%\begin{itemize}
%\item UQ is important for many fields
%\item As in previous work, uncertainties come from four sources:
%\begin{enumerate}
%\item Effective potentials (optical potentials)
%\item Approximations made to the few-body problem
%\item Structure functions
%\item Degrees of freedom left out of the model space
%\end{enumerate}
%\item Review where we were last time JPG 2014
%\item discuss progress we made on each of these in general terms
%\item outline of paper
%\end{itemize}x

Uncertainty quantification has gained a great deal of prominence in nuclear theory over the past several years.  From \emph{ab initio} methods to macroscopic theories, many groups are tackling the broad challenge of including meaningful uncertainties on their theoretical calculations \cite{Melendez2017,Furnstahl2015,Wesolowski2016,Schindler2009,Fukui2014,Schunck2015,Bernhard2016,Sangaline2016,Utama2016}.  Although standard $\chi^2$-minimization techniques and covariance matrix propagation had been the standard for decades, Bayesian methods have recently become the gold standard in uncertainty quantification, aimed at more reliably reporting calculated uncertainties, particularly from parameter optimization, e.g. \cite{Neufcourt2018}.  Building off of these improvements, there has also recently been a push to investigate techniques such as Gaussian Processes (GP) \cite{Novak2014,Neufcourt2018,Neufcourt2019} and other machine learning methods \cite{Neufcourt2018,Utama2016a,Lovell2020,Wang2019,Regnier2020} to create emulators in addition to quantifying uncertainties.   %\IR{Is there more that we need to say here?}

In few-body reaction theory, the typical means of quantifying uncertainties coming from approximations made to the theory was to compare a calculated observable (such as a differential cross section) to the same observable calculated using a more exact theory \cite{Nunes2011,Upadhyay2012,Deltuva2007,Capel2012,Titus2016}.  This comparison, however, only indicates the relative uncertainty between methods, not the absolute uncertainty on a calculation.  Parametric uncertainties were studied in much the same way - comparing calculations using different parameterizations.  However, these uncertainties can come from many sources.  The four main sources of uncertainty in few-body reaction theories are:
\begin{enumerate}
\item effective potentials (the optical potential)
\item approximations made to the few-body problem
\item the structure functions used
\item degrees of freedom left out of the model space
\end{enumerate}
\noindent Our group began to explore uncertainties within few-body reactions in the first special issue of this journal aimed at bridging the gap between experiment and theory \cite{Lovell2015}.  At that point, the state-of-the-art in the field, as mentioned previously, was using standard $\chi^2$ minimization to optimize the optical potential parameters fit elastic scattering data.  Subsequently, different parameterizations or models were compared based on relative differences.  Under this assumption, we found only 10-30\% differences between various optical model parameterizations, even for reactions on highly exotic nuclei, such as $^{132}$Sn(d,p)$^{133}$Sn.  Not only was this a basic arbitrary procedure to estimate uncertainties in the optical model, it provided no avenue to quantify the uncertainties coming from items (ii)--(iv).  

In the meantime, the many studies on uncertainty quantification for few-body reactions have greatly improved our understanding and helped establish the Bayesian framework in this area.  This is in line with improvements made in the broader nuclear physics community.  We will now briefly discuss each of the new developments and provide more detail in the next sections of this paper.  First, with regards to (i) the optical potential, we have replaced the frequentist propagation of uncertainties \cite{Lovell2017} with a full Bayesian analysis \cite{Lovell2018}. This decision was strongly motivated by the direct detailed comparison of the frequentist and Bayesian methods for reactions on stable targets \cite{King2019}.  For (ii) the approximations made to the few-body problem, we have compared the distorted-wave Born approximation (DWBA) and the adiabatic wave approximation (ADWA) using both the frequentist \cite{King2018} and Bayesian \cite{Lovell2018} optimizations.  Although this comparison does not take into account the model defects in either approximation, we are now able to determine whether these two models are consistent within the uncertainties propagated from the optical model.  With regards to (iii) and (iv), small studies have been started to incorporate these uncertainties as well, particularly focusing on the single-particle potential in transfer reactions and analyzing the change in the uncertainties when a more exact few-body model is used.  %In a full Bayesian manner, we are also now able to constrain (iii) \IR{the structure function to the asymptotic normalization coefficient (ANC) which will be presented for the first time in this work.}  We can also study (iv) the effect of the degrees of freedom left out of the model space, specifically using the coupled channel born approximation (CCBA) to couple excited states to the ground state - which was tangentially explored in a frequentist frame in \cite{Lovell2017}.

In this paper, we highlight the improvements that have been made to uncertainty quantification of few-body models with regards to the above four sources.  First, we outline the physical problem, the difference in philosophy between the frequentist and Bayesian methods in our context, and the various reaction models that are used throughout this work in Section \ref{sec:stats}.  In Section \ref{sec:optical}, we discuss the improvements that have been made to quantifying uncertainties coming from the parametrization of the optical model potential; in Section \ref{sec:fewbody}, we show quantified differences in approximations made to the few-body problem; in Section \ref{sec:structure}, we discuss the near-term progress that can be made on the structure functions and adding degrees of freedom to the model space that had previously been removed - both now in a Bayesian framework.
%in Section \ref{sec:structure}, \IR{we show improvements to the description of the structure functions in single-nucleon transfer reactions}; and in Section \ref{sec:dof}, we describe uncertainties on the effect of degrees of freedom left out of the model space.  
Finally, we conclude and give broad remarks about the next steps for uncertainty quantification in few-body reaction theory in Section \ref{sec:outlook}.  

\section{Reaction theory and statistical considerations}
\label{sec:stats}

%\begin{itemize}
%\item Explain the philosophy:  using experimental data where available
%\item what are the parameters, what are the models and what are the data?
%\item General discussion of the sorts of observables that we may have at hand.
%\item Frequentist philosophy and methods
%\item Bayesian philosophy and methods
%\end{itemize}

We focus on two methods of quantifying uncertainties which have seen a great deal of use in nuclear physics both historically and recently:  standard $\chi^2$ minimization and covariance matrix propagation - referred to here as frequentist methods - and Bayesian methods.  In both cases, the main idea is to fit a theoretical model, $\sigma ^\mathrm{th}(\mathbf{x})$ with free parameters $\mathbf{x}$, to a set of experimental data, $\sigma_i^\mathrm{exp}$ with experimental errors $\Delta \sigma _i$.  

The parameters, $\mathbf{x}$, that we aim to optimize are those of the optical model, an effective potential describing the interaction between a light projectile and a heavy target.  The optical potentials typically have real and imaginary terms,
\begin{equation}
U(r) = V(r) + i W(r) + iW_s(R),
\label{eq:pot}
\end{equation}

\noindent  where the imaginary term takes into account the loss of flux to reaction channels not explicitly included in the model.  The potentials contain three parts, a volume term, surface term, and spin orbit term (in addition to the standard Coulomb term, e.g. \cite{Fukui2014}, when charged projectiles are considered).  In the reaction models considered here, the volume term contains a real and imaginary part parameterized as a Woods-Saxon,
\begin{equation}
V(r) = -\frac{V_o}{1+\exp((r-R_o)/a_o)},
\end{equation}

\noindent and
\begin{equation}
W(r) = -\frac{W_o}{1+\exp((r-R_w)/a_w)},
\end{equation}

\noindent where $V_o$, $R_o$, and $a_o$ ($W$, $R_w$, and $a_w$) are the depth, radius, and diffuseness of the real (imaginary) term of the potential.  The surface term, $W_s(R)$, typically contains only an imaginary term which is parametrized as the derivative of a Woods-Saxon (parameters $W_s$, $R_s$ and $a_s$).  The spin-orbit term is also parametrized as a derivative of a Woods-Saxon, but these parameters are typically held constant, because unpolarized elastic scattering is not very sensitive to this piece of the interaction.  In total, we have nine free parameters (in each case, $R_i=r_iA^{1/3}$ where $A$ is the mass of the target nucleus and $r_i$ is the fitted parameter).

These parameters have historically been fitted separately for elastic scattering of incident neutrons and protons as a function of the mass and charge of the target, and the incident particle energy, e.g. \cite{Becchetti1969,Koning2003}.  
Such global parameterizations are able to provide a fair description overall, but when considering one single data set, for a given target and a given beam energy, it is common to improve upon the description obtained with global potentials.
%Often, these global parametrization will be further optimized to the specific reaction of study.  
This is the methodology that we employ in the current study:  we begin with a global optical model parametrization (typically the Bechetti and Greenlees (BG) \cite{Becchetti1969}) for the initialization of the minimization procedure and optimize the parameters with respect to a single reaction.  We focus primarily on elastic scattering data, $d\sigma/d\Omega(\theta)$ but also discuss the effects of including total or reaction cross sections, $\sigma _\mathrm{tot}$ (for neutron scattering) or $\sigma_R$ (for proton scattering), and vector analyzing powers, $Re(iT_{11}(\theta))$.  Note that these observables are sometimes included in fitting global optical potentials (e.g. \cite{Koning2003}).  We use experimental data where available; Table \ref{tab:expdata} contains the types of reactions, the beam energies, and the original references for the experimental data used in the rest of this work.  

\begin{table}
\centering
\begin{tabular}{l|c|c|c}
\textbf{Reaction} & \textbf{Data type} & \textbf{Energy (MeV)} & \textbf{Citation}  \\ \hline
$^{40}$Ca(n,n)$^{40}$Ca & $d\sigma/d\Omega(\theta)$ & 11.9 & \cite{Tornow1982} \\
$^{40}$Ca(n,n)$^{40}$Ca & $d\sigma/d\Omega(\theta)$ & 13.9 & \cite{Tornow1982} \\
$^{40}$Ca(n,n)$^{40}$Ca & $Re(iT_{11}(\theta))$ & 13.9 & \cite{Tornow1982} \\
$^{40}$Ca(n,n)$^{40}$Ca & $\sigma _\mathrm{tot}$ & 14.1 & \cite{Mcdonald1964} \\
$^{40}$Ca(p,p)$^{40}$Ca & $d\sigma/d\Omega(\theta)$ & 12.5 & \cite{Aoki1996} \\
$^{40}$Ca(p,p)$^{40}$Ca & $d\sigma/d\Omega(\theta)$ & 14.5 & \cite{Aoki1996} \\
$^{40}$Ca(p,p)$^{40}$Ca & $Re(iT_{11}(\theta))$ & 14.5 & \cite{Aoki1996} \\
$^{40}$Ca(p,p)$^{40}$Ca & $\sigma _R$ & 14.48 & \cite{Dicello1970} \\
$^{40}$Ca(p,p)$^{40}$Ca & $d\sigma/d\Omega(\theta)$ & 26.3 & \cite{Mccamis1986} \\
$^{40}$Ca(p,p)$^{40}$Ca & $d\sigma/d\Omega(\theta)$ & 27.5 & \cite{Mccamis1986} \\
$^{40}$Ca(p,p)$^{40}$Ca & $Re(iT_{11}(\theta))$ & 26.3 & \cite{Mccamis1986} \\
$^{40}$Ca(p,p)$^{40}$Ca & $\sigma _R$ & 24.5 & \cite{Carlson1975} \\
$^{40}$Ca(d,d)$^{40}$Ca & $d\sigma/d\Omega(\theta)$ & 30.0 & \cite{Roche1974} \\ 
%\IR{$^{40}$Ca(d,d)$^{40}$Ca$(0_1^+,2_1^+)$ }& $d\sigma/d\Omega(\theta)$ & 30.0 & \cite{Roche1974} \\ 
%$^{40}$Ca(d,p)$^{41}$Ca(g.s.) & $d\sigma/d\Omega (\theta)$ & \IR{???} & \IR{No data here} \\
\end{tabular}
\caption{Overview of the experimental data used in this work.  First columns gives the reaction at the energy listed in the third column.  The second column lists the type of data:  $d\sigma/d\Omega (\theta)$ for the differential cross section, $Re(iT_{11}(\theta))$ the analyzing powers (polarization data), and $\sigma _{R/tot}$ for either the reaction cross section or total cross section.  A reference for each data set is listed in the fourth column.  
%\IR{Going to need some way to distinguish the inelastic cross sections, e.g. $2^+$ }
}
\label{tab:expdata}
\end{table}

%%%%%%%%%%%%%%%%%%%%%%%%%%%%%%%%%%%%%%%%%%%%%%%%%
\subsection{Reaction models}
\label{sec:theory}

There are several reaction models that we consider in this work, all of which are not computationally demanding. 
We should emphasize that although there are several reaction models of higher complexity (e.g. \cite{xcdcc-summers,xcdcc-summers2,xcdcc-moro-prl,deltuva2009b}) and a number of frameworks for the optical potential that go well beyond the simple parameterization introduced in Eq.(\ref{eq:pot}) (e.g. \cite{non-local,dom,rotureau2018}), it is critical to initially explore the new statistical tools with simpler models that enable a full investigation of this statistical methodology.

For most of the cases that we study, we fit the optical potential parameters to reproduce single-channel elastic scattering, which is calculated using partial wave decomposition as explained in \cite{ThompsonNunes}. For each projectile-target angular momentum, the S-matrix $S_L$ (where $|S_L|^2$ is the probability that the projectile gets absorbed by the target) is obtained from solving the scattering equation and matching the solution to the known asymptotic behavior.  From there, all reaction observables can be calculated (for details see Section 3.2 of \cite{ThompsonNunes}).
%\begin{equation}
%\frac{d\sigma}{d\Omega}(\theta) = \frac{1}{4k^2} \left | \sum \limits _{L=0} ^\infty (2L+1)P_L(\mathrm{cos} \theta)  (S_L-1) \right |^2, 
%\label{eqn:el}
%\end{equation}
%
%\noindent where $k$ is the incoming moment, $P_L(\mathrm{cos}\theta)$ are the Legendre polynomials, and $S_L$ is the single-channel scattering matrix.  For complex potentials the total and reaction cross sections can also be defined through the S-matrix elements,
%\begin{equation}
%\sigma_{\mathrm{tot}/R} = \frac{\pi}{k^2} \sum \limits _{L=0} ^\infty (2L+1)(1-|\textbf{S}_L|^2),
%\end{equation}
%\noindent and the polarization can be defined as \IR{Also define the polarization.}  In this work, we use the vector analyzing powers, $Re(iT_{11}) = \sqrt{3}/2 A_y(\theta)$.

%To include inelastic scattering information, we use the coupled-channels Born approximation (CCBA).  This formulation couples the inelastic channels to the elastic channel by solving $N$ coupled equations
%\begin{equation}
%\left [ H _\alpha - E_\alpha \right ] \psi _{\alpha \alpha _{i}} (R_\alpha) + \sum \limits _{\beta \neq \alpha} ^N V_{\alpha \beta} \psi _{\alpha \beta} (R_\beta) =0
%\end{equation} 

%\noindent Here, $\alpha_i$ denotes the incoming channel, $\psi _{\alpha \alpha _{i}}$ is the wavefunction for the incoming channel, and $V_{\alpha \beta}$ is the coupling potential \cite{ThompsonNunes}.

Here we consider two models to predict the single-neutron transfer cross section for $A(d,p)B$, namely the one-step distorted-wave Born approximation (DWBA) and the adiabatic-wave approximation (ADWA).  In DWBA, the elastic scattering of the incoming deuteron is described by an effective deuteron-target interaction, $V_{dA}$.  Instead of solving the full three-body problem, the true three-body wave function is replaced by the elastic channel (represented by the deuteron distorted wave multiplied by the corresponding bound state of the deuteron) \cite{ThompsonNunes}:
\begin{equation}
\textbf{T} ^\mathrm{DWBA} _\mathrm{post} = \langle \Phi_{nA} (\vec{r}_{nA}) \chi_p (\vec{R}_f) | V_{np}+\Delta | \Phi _{np}(\vec{r}_{np}) \chi _{d\vec{k}_{i}}(\vec{R}_i) \rangle,
\label{eqn:TDWBA}
\end{equation}

\noindent with $\Phi_{np}(\vec{r}_{np})$ the bound-state deuteron wave function, $\Phi_{nA} (\vec{r}_{nA})$ the final bound-state wave function of $B$, $\chi _{d\vec{k}_{i}}(\vec{R}_i)$ the distorted wave of the $d+A$ system, $\chi_p (\vec{R}_f)$ the distorted wave of the proton interacting with $B$, $V_{np}$ the deuteron binding potential, and $\Delta$  the difference between the $A+p$ and $B+p$ optical potentials.  Note that the coordinates introduced in Eq. (\ref{eqn:TDWBA}) are the standard Jacobin coordinates in the entrance ($\vec{r}_{np},\vec{R}_i$) and exit ($\vec{r}_{nA},\vec{R}_f$) channels as in Ref.\cite{ThompsonNunes}.

However, because of its small binding energy, it is likely that deuteron breakup will occur in the field of the target, and this has been shown to influence other channels such as the transfer \cite{Nunes2011}.  For this reason, many reaction theories start from the three-body $n+p+A$ scattering problem, and solve with different levels of approximations.

Since solving the three-body scattering problem exactly (in a Faddeev formalism) presents challenges and requires significant computational time, it is not feasible to use these methods in the context of the Bayesian approach discussed in Section \ref{sec:bayes}.  Instead we use the adiabatic approximation of Johnson and Tandy \cite{tandy} by considering that the excitation energy of the deuteron is negligible compared to the beam energy.  This approximation leads to a simplified three-body equation:
\begin{equation}
\left [ T_R +V_{pA}(\vec{r},\vec{R}) +V_{nA}(\vec{r},\vec{R}) -(E-\epsilon_0) \right ] \Psi ^\mathrm{ad} (\vec{r},\vec{R}) = 0,
\label{eqn:ad}
\end{equation}

\noindent where $T_R$ is the center of mass kinetic energy, $V_{nA}$ and $V_{pA}$ are the neutron-target and proton-target optical potentials, $E$ is the incident beam energy, and $\epsilon_0$ is the binding energy of the deuteron.  
Furthermore, by considering a Weinberg expansion for $\Psi ^\mathrm{ad} (\vec{r},\vec{R})$, it is possible to integrate out the $\vec r$ variable in Eq. (\ref{eqn:ad}) and obtain a single-channel scattering equation, greatly reducing the computational time \cite{tandy}. The adiabatic wave function is then used in the post-form transfer T-matrix,

\begin{equation}
\textbf{T}^\mathrm{ADWA} _\mathrm{post} = \langle \Phi _{nA} (\vec{r}_{nA}) \chi _p (\vec{R}_f) |V_{np} | \Psi ^\mathrm{ad} (\vec{r}, \vec{R}) \rangle.
\label{eqn:TADWA}
\end{equation}
This approach is usually referred to as the finite-range adiabatic wave approximation (ADWA).

In both DWBA or ADWA, the transfer cross section is obtained from the norm squared of the T-matrix.
%Eq. (\ref{eqn:TDWBA}) or (\ref{eqn:TADWA}).

%%%%%%%%%%%%%%%%%%%%%%%%%%%%%%%%%%%%%%%%%%%%%%%%%
\subsection{Frequentist methods}
\label{sec:freq}

In this work, we refer to standard $\chi^2$ minimization and propagation of the resulting covariance matrix as frequentist methods.  This type of optimization has been the standard in the field for many decades.  The goal is to describe a true function $\sigma (\theta)$ with a model $\sigma (\textbf{x},\theta)$ that has $N$ free parameters, $x_1,x_2,...x_N$.  To constrain these parameters, we have a set of M experimental data pairs, $\{ (\theta_1, \sigma _1), (\theta_2, \sigma _2) ... (\theta_M, \sigma _M)\}$, each of which has an associated experimental uncertainty, $\Delta \sigma _i$.  The typical assumption is that the experimental values are uncorrelated with one another,
\begin{equation}
\sigma ^\mathrm{exp} _i = \sigma (\theta _i) + \epsilon_i,
\end{equation}

\noindent where each $\epsilon _i$ is normally distributed,
\begin{equation}
\epsilon _i \sim \mathcal{N}(0,(\Delta \sigma _i)^2).
\end{equation}

\noindent In matrix form, this is written, $\sigma ^\mathrm{exp} \sim \mathcal{N}(\sigma, \Sigma)$, where $\Sigma$ is an $M \times M$ matrix with $(\Delta \sigma_i)^2$ on the diagonal.  The residuals between the experimental values and the theory predictions, $\sigma_i^\mathrm{exp} - \sigma (\textbf{x},\theta_i)$, are also normally distributed as $ \mathcal{N}(0,\Sigma)$.  Maximizing the likelihood function for $\textbf{x}$ is equivalent to minimizing the corresponding objective function,
\begin{equation}
\chi^2_{UC} =  \sum \limits _{i=1} ^{M} \frac{(\sigma^\mathrm{exp} _i - \sigma(\mathbf{x},\theta_i))^2}{(\Delta \sigma_i) ^2},
\label{eqn:chiUC}
\end{equation}

\noindent where $UC$ stands for uncorrelated.  Equation (\ref{eqn:chiUC}) is proportional to the standard $\chi^2$ function, and its minimization leads to the determination of a best fit set of parameters, $\hat{x}_{UC}$.  The 95\% confidence intervals around $\sigma(\hat{x}_{UC},\theta)$ can be constructed by assuming that the true parameter values are distributed normally around the best-fit parameter set, so parameters can be drawn from 
\begin{equation}
\mathcal{N}(\hat{x},\mathbb{C}_p) \sim \exp [ (\textbf{x} - \hat{x}_{UC} ) ^T \mathbb{C}_p (\textbf{x} - \hat{x}_{UC}) ] 
\label{eqn:parmdraws}
\end{equation}

\noindent where $\mathbb{C}_p$ is the $N \times N$ parameter covariance matrix.  The goodness of fit is taken into account by scaling this covariance matrix by 
\begin{equation}
s^2 = \frac{\chi^2_{UC}}{M-N}.
\end{equation}

\noindent Then $\mathbb{C}_p$ in Eq. (\ref{eqn:parmdraws}) is replaced by $s^2\mathbb{C}_p$.

However, if we consider differential elastic cross sections as calculated in the standard manner using a partial wave decomposition, where the use of the Legendre polynomials inherently correlates the cross section at each angle, we can explore a different approach to include these correlations in the fitting procedure.  We can then introduce a correlated $\chi^2$ function:

\begin{equation}
\chi^2_C = \sum \limits _{i=1} ^{M} \sum \limits _{j=1} ^{M}  \mathbb{W}_{ij} (\sigma^\mathrm{exp} _i - \sigma(\mathbf{x},\theta_i))(\sigma^\mathrm{exp} _j - \sigma(\mathbf{x},\theta_j)),
\label{eqn:chiC}
\end{equation}

\noindent where $\mathbb{W}_{ij}$ are the $(ij)^\mathrm{th}$ elements of the inverse of the data covariance matrix defined as 
\begin{equation}
\mathbb{W} = (\mathbb{C}_m + \Sigma)^{-1},
\label{eqn:expCovariance}
\end{equation}

\noindent with $\mathbb{C}_m$ being the model covariance matrix between the angles of the experimental data points and $\Sigma_{ii}=(\Delta \sigma_i)^2$.  To calculate the model covariance matrix, parameter sets in the optical model are randomly sampled around the initial parametrization and run through the model.  $\mathbb{C}_m$ is then explicitly calculated as the covariance between each of the angles included in the fitting procedure.  The elements of $\mathbb{C}_m$ do not have to be positive, leading to interference between the different angles in the model, and therefore, $\chi^2_C < \chi^2_{UC}$.  In addition, because the model correlation matrix is not normalized,  $\chi^2/M \leq 1$ is no longer the definition of a good statistical fit.

To construct the 95\% confidence intervals, once the best-fit set of parameters, $\hat{x}_{UC}$ or $\hat{x}_C$, is determined, parameter sets are sampled from Eq. (\ref{eqn:parmdraws}), where $\mathbb{C}_p$ and $s^2$ are determined from either the uncorrelated or correlated $\chi^2$, and run through the model.  At each evaluated angle, the top 2.5\% and bottom 2.5\% of the calculations are removed to leave 95\% intervals.  

%%%%%%%%%%%%%%%%%%%%%%%%%%%%%%%%%%%%%%%%%%%%%%%%%
\subsection{Bayesian methods}
\label{sec:bayes}

In contrast to frequentist methods, where the fundamental interpretation is that out of a list of options, one must occur, Bayesian statistics give the probability of a single occurrence independent of all others.  In addition, with Bayesian methods, we are able to introduce prior information in the statistical formulation, compare two theories, or even mix models, all within a consistent framework.  We give a brief overview in the following section, but more details can be found in \cite{BayesBook,Trotta2008}.

For a hypothesis, $H$, (in this work, a given model with some set of free parameters) that is constrained by some experimental data, $D$, Bayes' theorem is written as 
\begin{equation}
p(H|D) = \frac{p(H)p(D|H)}{P(D)},
\label{eqn:bayes}
\end{equation}

\noindent where the prior, $p(H)$, is information that is known about the model before seeing the experimental data, and the likelihood, $p(D|H)$ contains information about the goodness of fit between the model and the data, here a standard normal distribution, $\exp ^{-\chi^2/2}$.  For our $\chi^2$ function, we only consider the uncorrelated $\chi^2$ of Eq. (\ref{eqn:chiUC}).  Bayes' theorem allows for the calculation of the posterior distribution, $p(H|D)$, which is the most likely probability distribution of the fitting parameters conditional on the experimental data.  The last piece of Eq. (\ref{eqn:bayes}) is $p(D)$, the Bayesian evidence which typically consists of a weighted sum over all possible hypotheses - or models.  

Often, the Bayesian evidence is difficult or nearly impossible to calculate directly, so Monte Carlo methods are used to sample the posterior distribution directly.  Here, we use the Metropolis-Hastings Markov Chain Monte Carlo (MCMC) \cite{Metropolis1953,Hastings1970}.  We begin with an initial set of optical model parameters $\textbf{x}_i$ from which the prior, $p(H_i)$, and likelihood, $p(D|H_i)$, are calculated.  A new set of parameters is sampled from a normal distribution, $\textbf{x}_f \sim \mathcal{N}(\textbf{x}_i,\epsilon \textbf{x}_0)$, where $\epsilon$ is a scaling factor that controls the step size through parameter space.  From the updated parameter space, $\textbf{x}_f$, the new prior and likelihood are calculated, $p(H_f)$ and $p(D|H_f)$.  This new parameter set is accepted if the following condition is fulfilled:
\begin{equation}
\frac{p(H_f)p(D|H_f)}{p(H_i)p(D|H_i)} > R,
\label{eqn:Rcondition}
\end{equation}

\noindent where $R$ is a random number between 0 and 1.  If $\textbf{x}_f$ is accepted, it becomes the initial parameter set, and a new random set of parameters is drawn.  If Eq. (\ref{eqn:Rcondition}) is not fulfilled, the parameter set $\textbf{x}_f$ is rejected and a new parameter set is drawn from $\mathcal{N}(\textbf{x}_i,\epsilon \textbf{x}_0)$.  This process continues until a predetermined number of parameter sets have been accepted.

Initially, there is no guarantee that the initial parameter set is within the targeted posterior distribution.  For this reason, we need a burn-in period where a certain number of parameter sets, $N_\mathrm{burn-in}$, are discarded, regardless of whether or not these sets were accepted by the Monte Carlo condition of Eq. (\ref{eqn:Rcondition}).  The end of the burn-in is signified by a likelihood and parameter distributions that oscillate around a mean value and are not consistently increasing or decreasing.  After the burn-in period, each accepted parameter set is directly correlated to the parameter sets accepted before it.  To remove this dependency, we do not record a certain number of accepted parameters, $N_\mathrm{thin}$ between each recorded parameter set.

Confidence intervals in Bayesian statistics are calculated slightly differently than confidence bands in the frequentist method.  In this case, 95\% confidence intervals are defined by the smallest interval, $[a,b]$, where
\begin{equation}
\int \limits _a ^b p(H_i|D) dx_i = 0.95
\label{eqn:bayesianCI}
\end{equation}

\noindent for a given variable $x_i$.  In practice, because our parameters are sampled numerically from the MCMC, the integral becomes a sum.  These intervals define the region where we believe, with a 95\% probability, that the true value of the cross section or optical potential parameters fall.  

%\IR{Should we discuss evidence?}

%%%%%%%%%%%%%%%%%%%%%%%%%%%%%%%%%%%%%%%%%%%%%%%%%
\subsection{Numerical details}
\label{sec:num}

%\begin{itemize}
%\item Details about the initialization of the UC and C fitting
%\item Details about the prior distributions
%\item Reference Table \ref{tab:expdata} where all of the data are
%\item Mention which reactions are needed for DWBA vs ADWA
%\item Cite fresco and NLAT - and that we have updates to these codes
%\end{itemize}

There are several optical potentials that are needed to calculate the single-nucleon transfer cross sections using DWBA or ADWA.  In this work, we  focused on calculating the $^{40}$Ca(d,p)$^{41}$Ca transfer cross section in the ground state (g.s.) at 28.4 MeV.  In DWBA, the deuteron-target interaction is needed at the incident deuteron energy.  For ADWA, the neutron-target and proton-target potentials are needed which we take at half the deuteron energy.
%and although there is some controversy over the energy of the neutron and proton, we choose these to be half of the incident deuteron energy.  
For both calculations, we additionally require the potential between the proton and the $^{41}$Ca at an energy $E-Q$, where $E$ is the incident deuteron energy and $Q$ is the $Q$-value of the reaction.  However, scattering data on $^{40}$Ca is more readily available than data on $^{41}$Ca - and there is very little difference between the optical potentials on these two targets - so we constrain this channel based on $^{40}$Ca-p scattering data, as in all of our previous studies \cite{Lovell2017,Lovell2018,King2018,King2019,CatacoraRios2019}.  These data are summarized in Table \ref{tab:expdata} in addition to listing the other data sets that we will later use to explore further constraints on the optical potential.  For all of the elastic scattering angular distribution data and total reaction cross sections, we take the experimental uncertainty to be 10\%.  In Section 3.4, we will also consider the vector analyzing powers. For those, we also take the uncertainty to be 10\% except in cases where the value of the analyzing power at a given angle is less than 5\% of the maximum value across all angles (e.g. when the analyzing power is very close to zero).  For these cases, we take the uncertainty to be 5\% of the maximum value. This provides a lower bound of the error, in line with real data.

For the uncorrelated frequentist calculations, we initialize each of the parameters with the Becchetti-Greenlees (BG) \cite{Becchetti1969} optical potential parameters for neutron and proton scattering and the An-Cai (AC) \cite{AC} optical potential for deuteron scattering.  With the $\chi^2$ minimization, it is possible that the parameters can fall into a minimum that is outside of the physical range of these parameters.  To prevent that, we fix some of the parameters during the optimization (typically those of the imaginary volume term).  The potential for the incoming proton at 14.5 MeV in the frequentist optimizations was initialized with the incoming neutron parameters to ensure that the geometry was similar for both neutrons and protons.

The Bayesian optimization is also initialized from the BG potential parameters.  In this case, we also must define a shape for the prior distributions.  As in our previous works, we take independent Gaussian priors for each of the optical potential parameters, centered at the Becchetti-Greenlees parameter values with a width of 100\% of those values.  This is a very wide prior, but it allows the data to drive the optimization, instead of being driven by the shape of the prior.  These assumptions about the shape of the prior were explored in \cite{Lovell2018}.

The frequentist minimizations for elastic scattering use \texttt{sfresco}, a best fit program using the \texttt{MINUIT} routines \cite{minuit}, packaged with \texttt{fresco} \cite{fresco}.  
The Bayesian statistical tools were developed from scratch as wrapper codes that call {\sc fresco}, {\sc sfresco} \cite{fresco} and {\sc nlat} \cite{nlat}. We will refer these wrappers collectively as {\sc QUILTR}, {\it \textbf{Q}uantified \textbf{U}ncertainties \textbf{I}n \textbf{L}ow-energy \textbf{T}heory for \textbf{R}eactions}. 
%The Bayesian optimization uses a modified version of \texttt{fresco}.  For both optimization techniques, the transfer reactions are calculated using a modified version of the code \texttt{NLAT} \cite{NLAT}.

%%%%%%%%%%%%%%%%%%%%%%%%%%%%%%%%%%%%%%%%%%%%%%%%%
\section{The optical potential}
\label{sec:optical} 

%\begin{itemize}
%\item Discuss frequentist uncorrelated and correlated chi2
%\item Discuss frequentist comparing models for transfer \IR{- maybe in the next section?}
%\item Highlight the limitations of the approach for UQ
%\item Discuss Frequentist vs Bayesian comparison (elastic/transfer)
%\item Constrained OM with elastic cross sections, reaction/total cross sections, and polarization data
%\item Other experimental conditions - angular range, multiple data sets
%\item Could try to evaluate/compare different minima (what does this mean?? in frequentist??) \IR{Perhaps evaluate the shapes of different minima?  Do these give the same elastic or transfer?  This may go more towards model averaging and a different study.}
%\item Still have not looked at whether constraining multiple reactions simultaneously reduces the uncertainties (e.g. include a mass/charge/energy dependence)
%\end{itemize}

Most of our efforts over the past several years have concentrated on constraining the parameters of the optical potential; we present a summary of the results of these studies here. Although the type of studies described in Sections 3.1-3.3 have been published in some form in recent years, in this work we bring all the features together, and apply these to a single new case (reactions on $^{40}$Ca) with consistent assumptions, both regarding the experimental data and the statistical methods.  

We use experimental data from three reactions, $^{40}$Ca(p,p), $^{40}$Ca(n,n), $^{40}$Ca(d,d) to constrain the optical potentials needed to calculate the single-neutron transfer cross section, $^{40}$Ca(d,p)$^{41}$Ca(g.s.) using both DWBA and ADWA with the data outline in Table \ref{tab:expdata}.  As most of the comparisons in this field has historically been performed using frequentist methods, we first compare the frequentist methods using the uncorrelated and correlated $\chi^2$ functions of Eqs. (\ref{eqn:chiUC}) and (\ref{eqn:chiC}).  

\subsection{Confidence intervals for angular distributions}

\begin{figure}
\centering
\begin{tabular}{cc}
\includegraphics[width=0.5\textwidth]{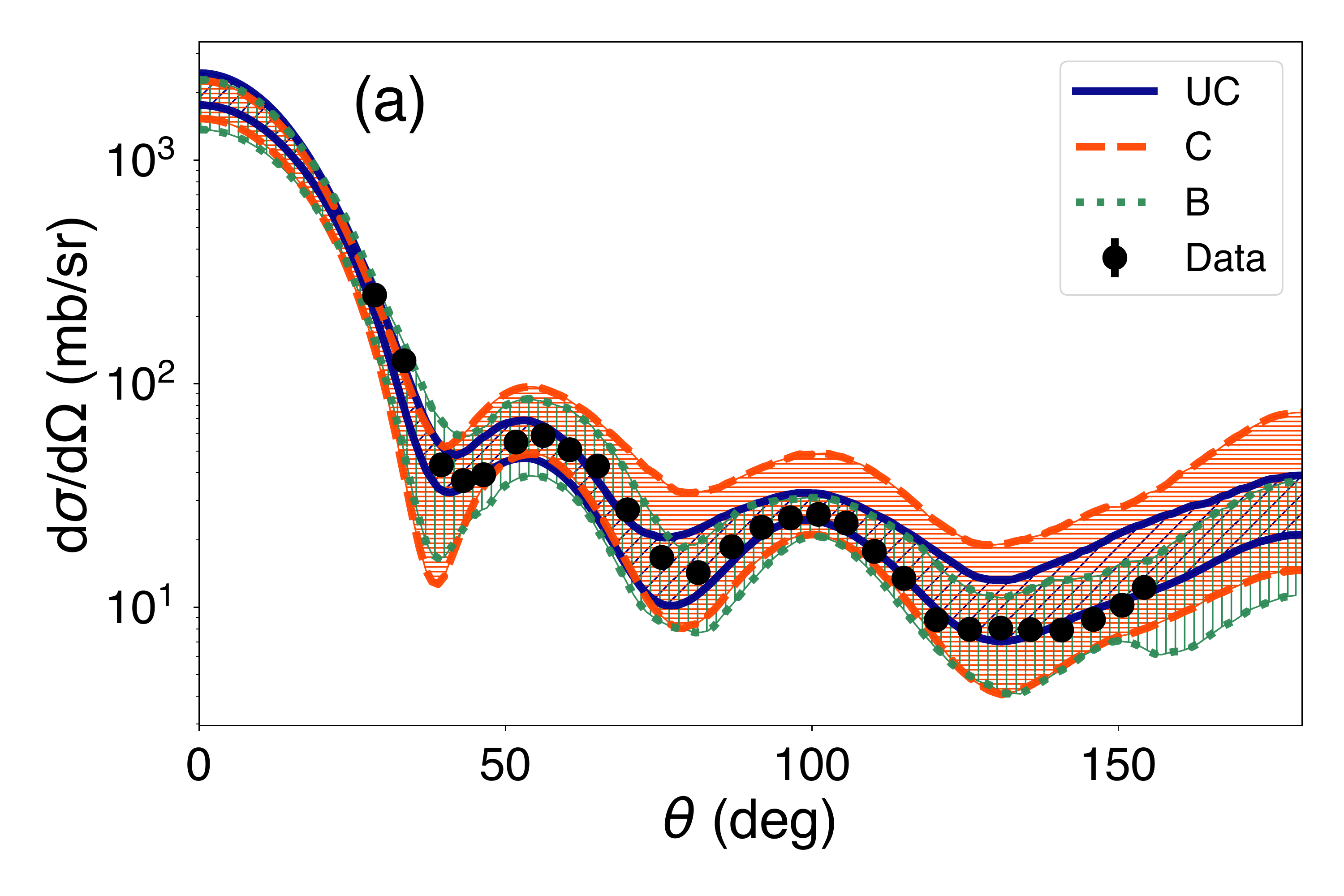} & \includegraphics[width=0.5\textwidth]{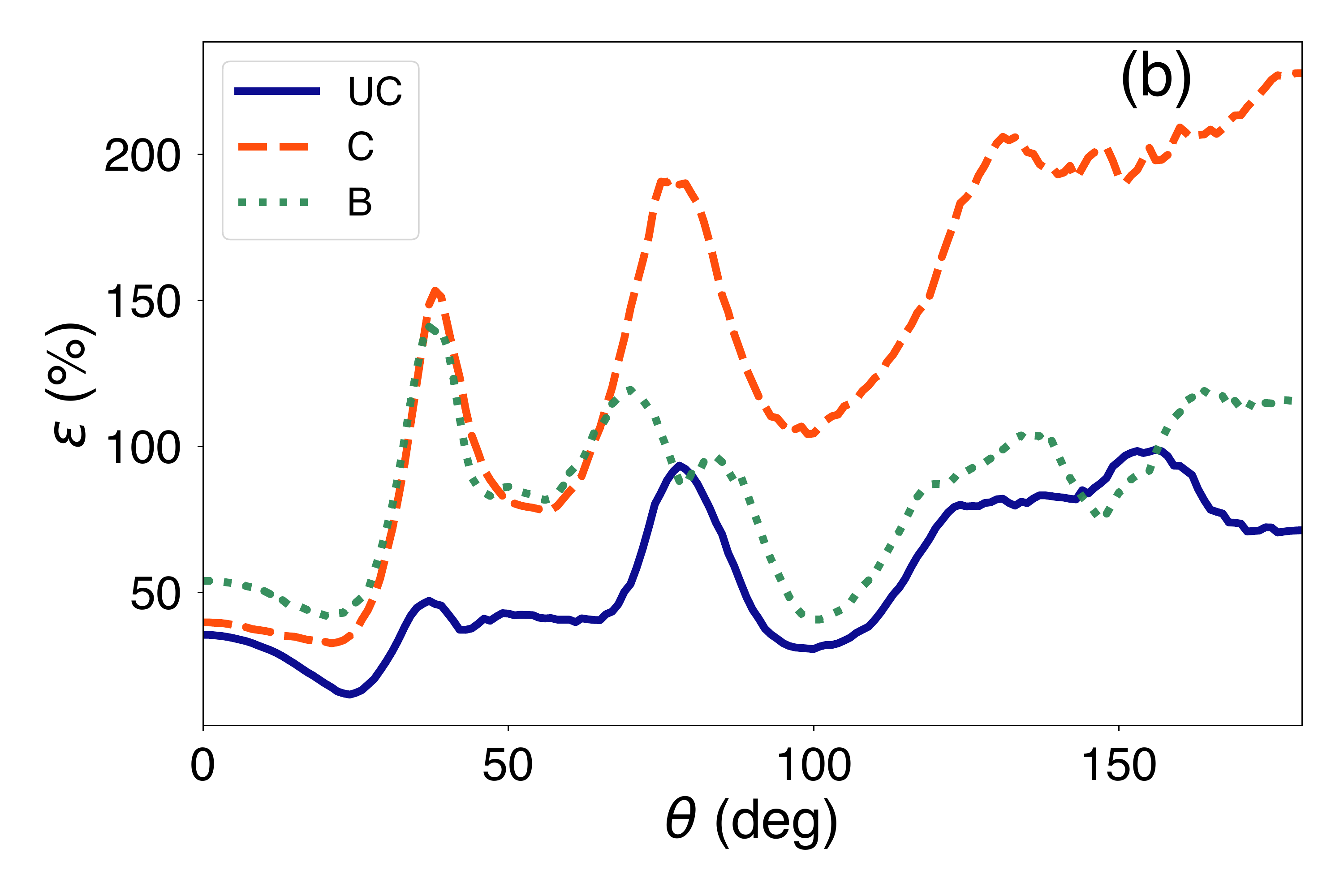} \\
\includegraphics[width=0.5\textwidth]{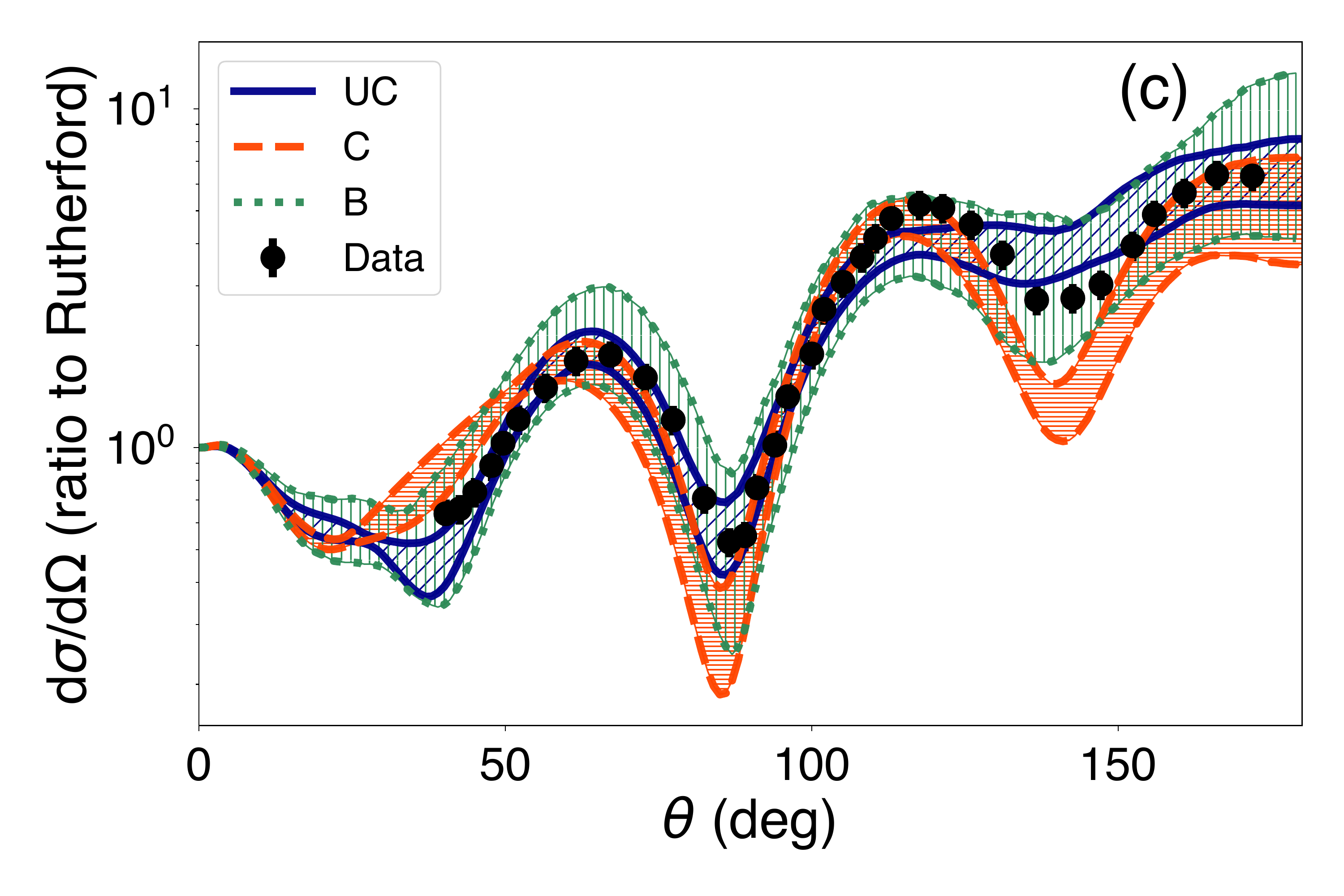} & \includegraphics[width=0.5\textwidth]{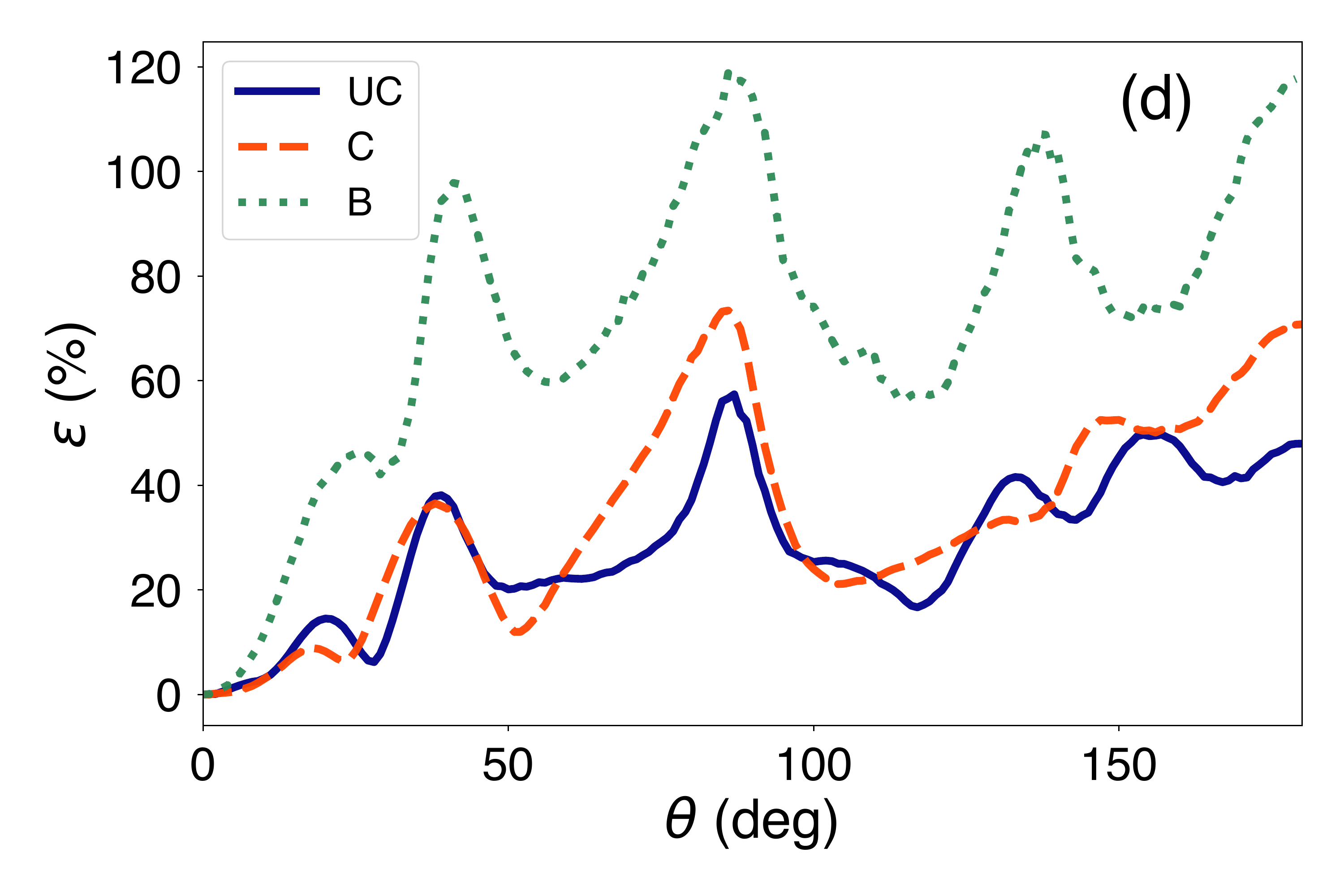} \\
\includegraphics[width=0.5\textwidth]{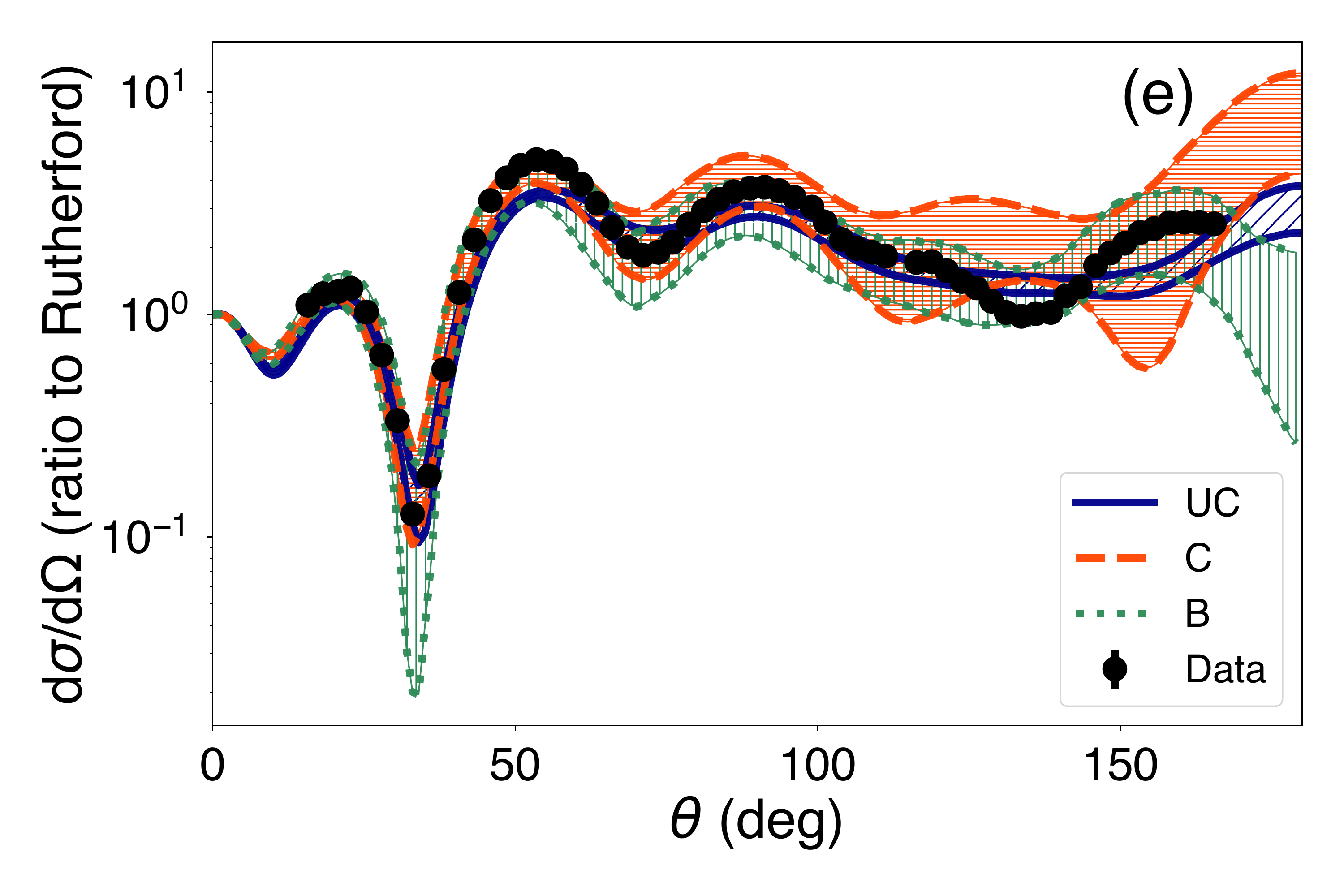} & \includegraphics[width=0.5\textwidth]{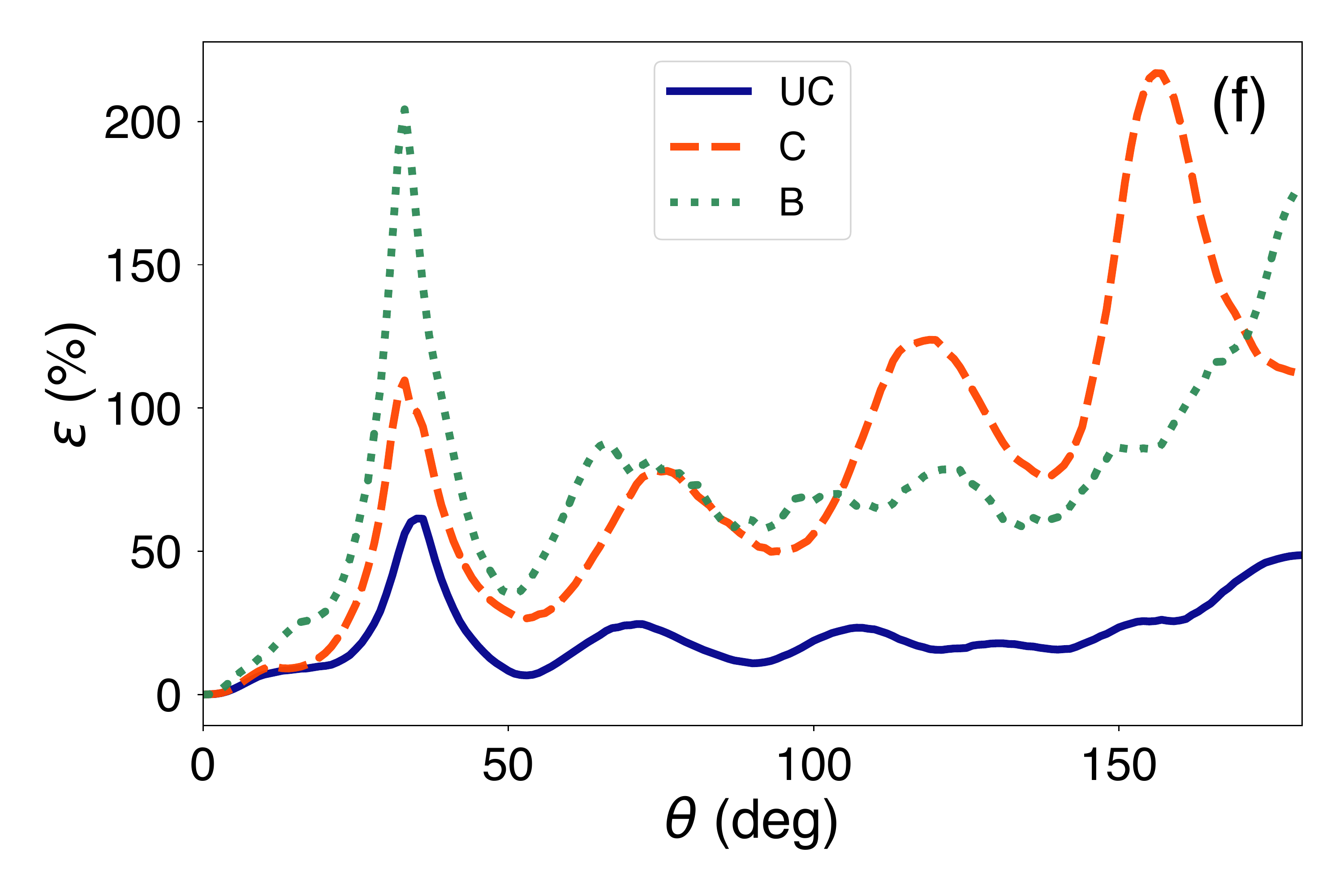} \\
\includegraphics[width=0.5\textwidth]{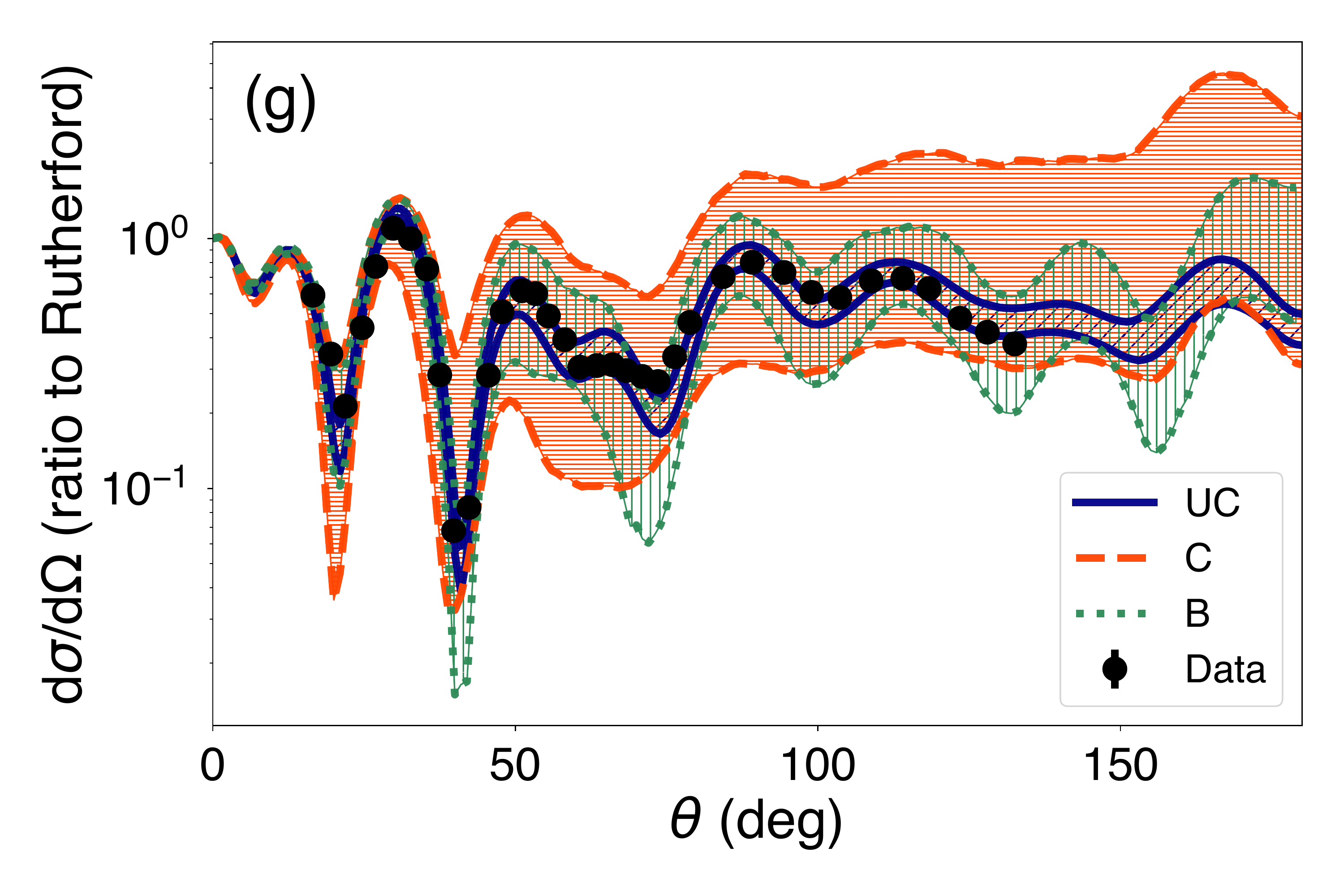} & \includegraphics[width=0.5\textwidth]{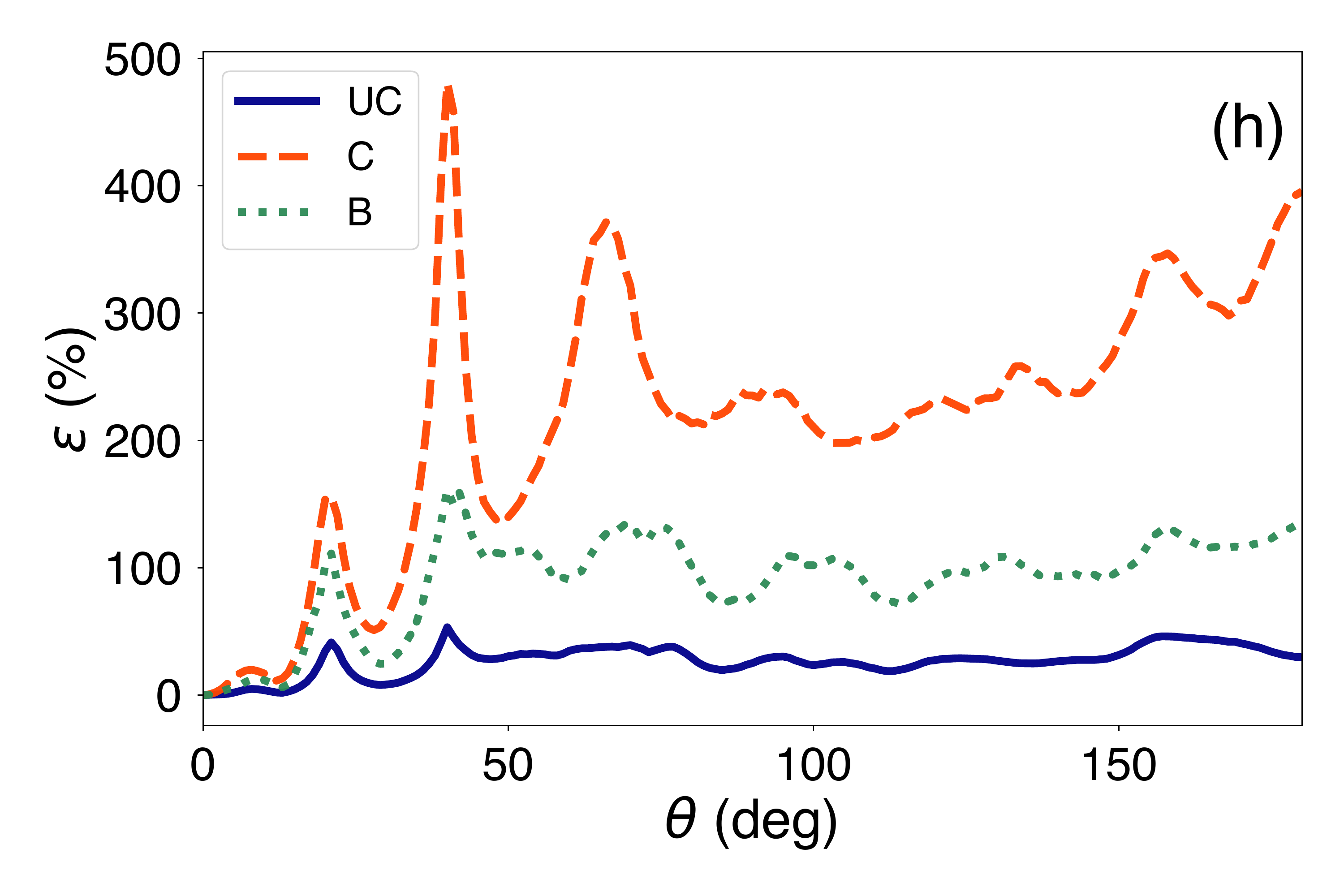} \\
\end{tabular}
\caption{95\% confidence bands (left column) and percentage uncertainty (right column) for $^{40}$Ca(n,n) at 13.9 MeV - (a) and (b) - $^{40}$Ca(p,p) at 14.5 MeV - (c) and (d) - $^{40}$Ca(p,p) at 26.3 MeV - (e) and (f) - and $^{40}$Ca(d,d) at 30.0 MeV - (g) and (h).  Comparison between the uncorrelated (\emph{UC}, in blue), correlated (\emph{C}, in red), and Bayesian (\emph{B}, in green) optimizations.}
\label{fig:UCCB}
\end{figure}

In Fig. \ref{fig:UCCB}, we show the 95\% confidence intervals for the uncorrelated (\emph{UC}, blue) and correlated (\emph{C}, red) fits for all incident and outgoing channels needed for the DWBA and ADWA calculations, (a) $^{40}$Ca(n,n) at 13.9 MeV, (c) $^{40}$Ca(p,p) at 14.5 MeV, (e) $^{40}$Ca(p,p) at 26.3 MeV, and (g) $^{40}$Ca(d,d) at 30.0 MeV.   In addition, in panels (b), (d), (f), and (g) we show the corresponding uncertainty on the differential cross sections defined as $\varepsilon = \Delta \sigma / \overline{\sigma}$, where $\Delta \sigma$ is the width of the 95\% confidence interval and $\overline{\sigma}$ is the best-fit cross section.
%\begin{equation}
%\varepsilon = \frac{\Delta{\sigma}}{\overline{\sigma}}
%\label{eqn:xsUncertainty}
%\end{equation}

We see, in all cases, the confidence intervals for the correlated fits are at least as large (but in most cases larger) than those for the uncorrelated fits.  This increase is seen even though the $\chi^2$ values are at least 5 times smaller in the correlated best fit compared to the uncorrelated best fit, due to the inclusion of the angular covariance matrix in the definition of $\chi^2_{C}$ in Eq. (\ref{eqn:chiC}).  The addition of this covariance matrix also allows the best fit the flexibility of not going through all of the data points, which is why we see the data outside of the confidence intervals, in particular for both proton-scattering reactions, in Fig. \ref{fig:UCCB} (c) and (e).  The standard, uncorrelated $\chi^2$ minimization requires that for every data point above the best fit, there is a data point below the best fit; this restriction is loosened when the angular covariance matrix is introduced in the correlated $\chi^2$ optimization.

\begin{figure}
\centering
\begin{tabular}{cc}
\includegraphics[width=0.5\textwidth]{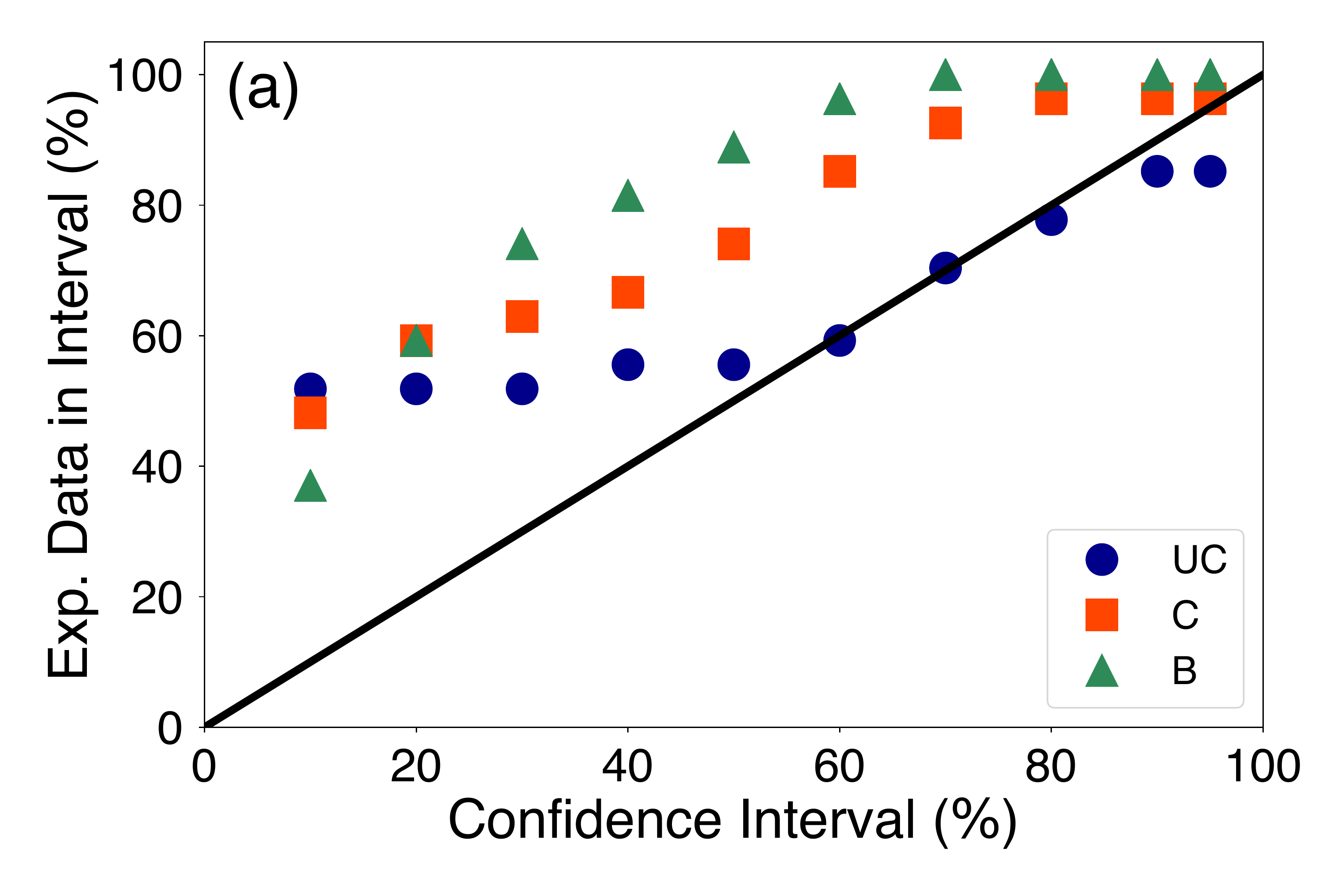} & \includegraphics[width=0.5\textwidth]{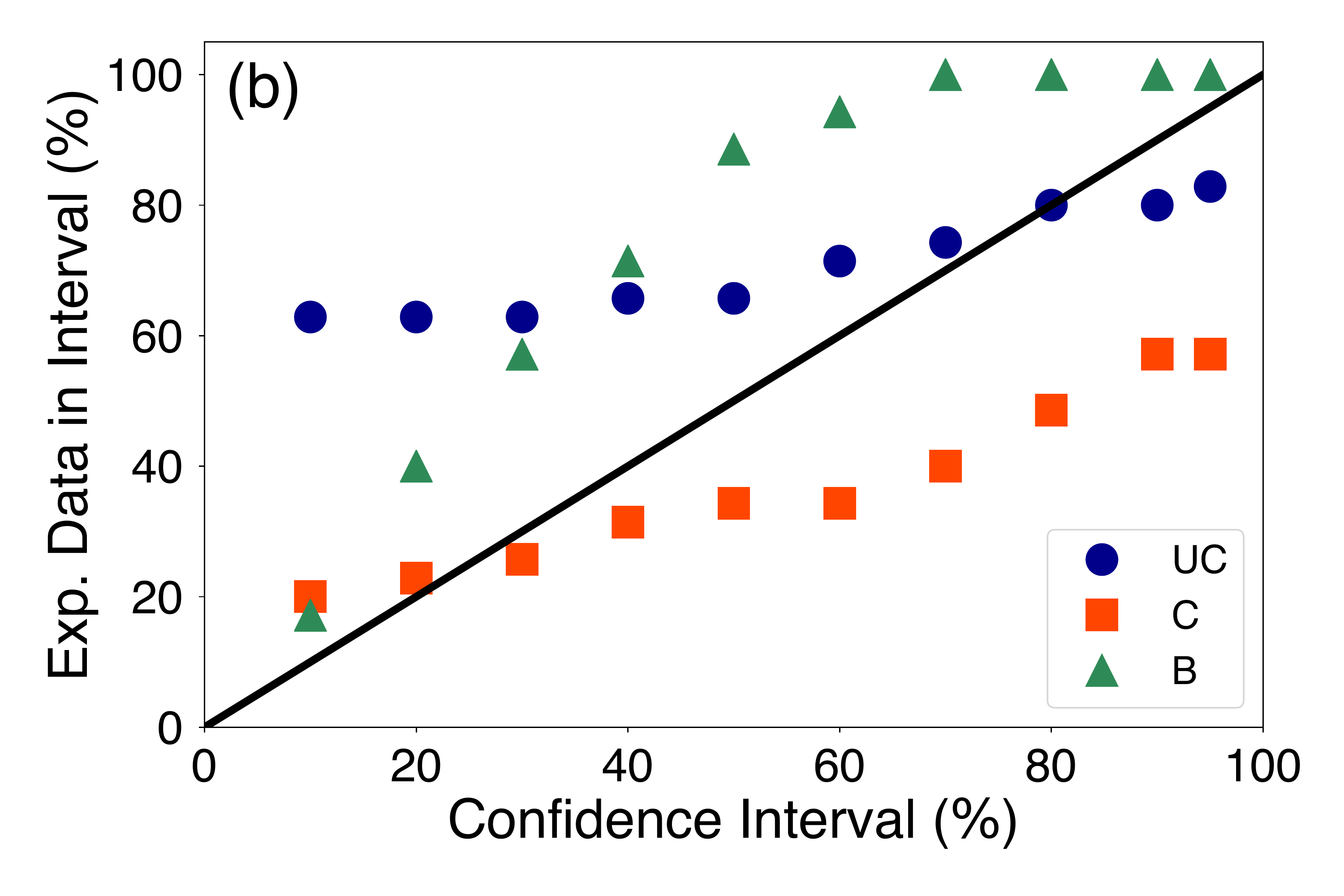} \\
\includegraphics[width=0.5\textwidth]{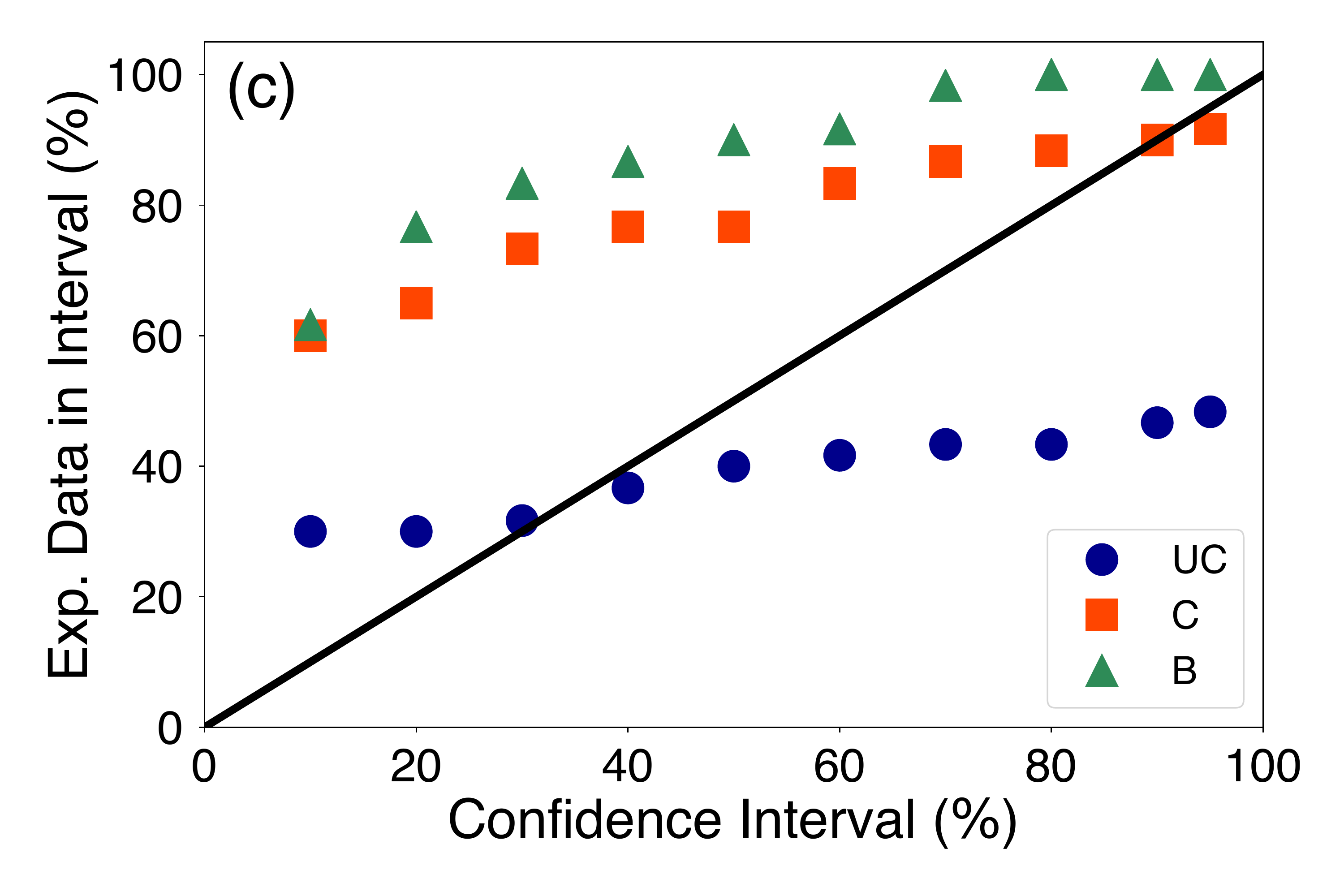} & \includegraphics[width=0.5\textwidth]{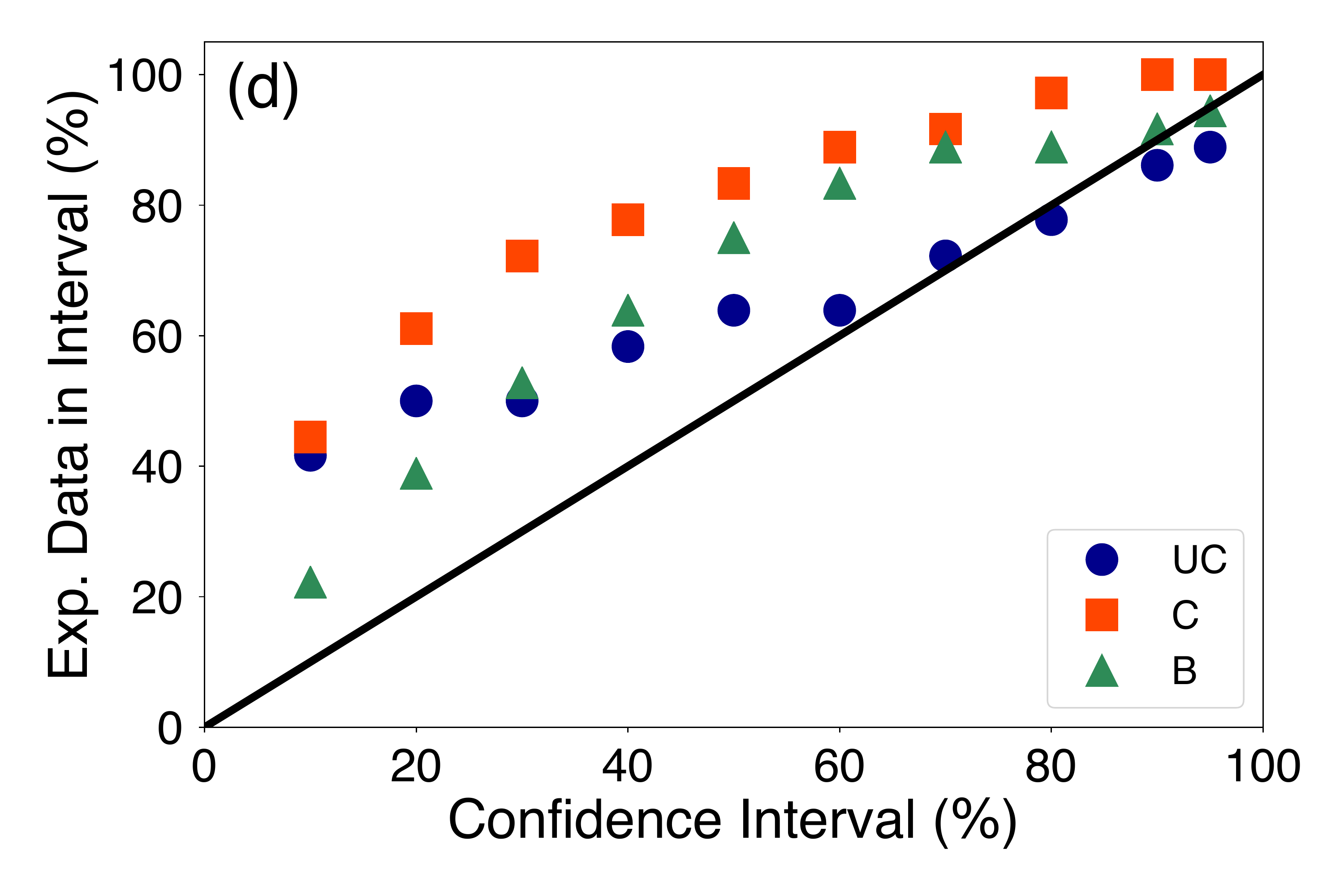} \\
\end{tabular}
\caption{Percentage of the the data that falls within a given confidence interval for the uncorrelated (UC, blue circles), correlated (C, red squares), and Bayesian (B, green triangles) minimizations, for the four elastic scattering reactions studied in this work, (a) $^{40}$Ca(n,n) at 13.9 MeV, (b) $^{40}$Ca(p,p) at 14.5 MeV, (c) $^{40}$Ca(p,p) at 26.3 MeV, and (d) $^{40}$Ca(d,d) at 30.0 MeV.}
\label{fig:percentages}
\end{figure}

In Fig. \ref{fig:percentages}, we calculate the percentage of the experimental data that falls within various confidence intervals for the elastic-scattering reactions shown in Fig. \ref{fig:UCCB}.  If the confidence intervals truthfully reproduced the uncertainties, X percent of the data would fall within the X percent confidence intervals, indicated by the solid black lines.  The uncorrelated confidence intervals consistently over-predict the uncertainties for small values of the confidence intervals, below $\sim 50$\%.  For the neutron and proton scattering, the trends in the uncorrelated optimization are flat compared to the solid black line.  The correlated optimization systematically over-predicts the uncertainty in each case except for proton scattering at 14.5 MeV, panel (b).  While over-predicting the uncertainty provides a more conservative estimate, the trends in Fig. \ref{fig:percentages} are not the same across the four scattering cases.%The uncorrelated confidence intervals, blue circles, almost consistently under-predict the true uncertainty (fall below the black line), particularly for the higher confidence intervals ($>50$\%).  In contrast, the correlated optimization over-predicts the uncertainty for  all cases but the proton scattering at 14.5 MeV, panel (b).  
%Although overestimating the uncertainty provides a more realistic value, by being more conservative, we see that the correlated confidence intervals, red squares, significantly over-predict the uncertainty.  

%\begin{table}
%\centering
%\begin{tabular}{c|ccc}
%\textbf{Reaction} & \textbf{Uncorrelated } & \textbf{Correlated} & \textbf{Bayesian} \\ \hline
%$^{40}$Ca(n,n)$^{40}$Ca at 13.9 & 88.9\% & 96.3\% & 100\% \\
%$^{40}$Ca(p,p)$^{40}$Ca  at 14.5 & 85.7\% & 57.1\% & 100\% \\
%$^{40}$Ca(p,p)$^{40}$Ca at 26.3 & 40.0\% & 100\% & 91.6\% \\
%$^{40}$Ca(d,d)$^{40}$Ca at 30.0 & 88.9\% & 90.0\% & 96.6\% \\
%\end{tabular}
%\caption{Percentage of the data that falls within each 95\% confidence interval for each reaction listed in column 1 (energy given in MeV).  This percentage is listed for the uncorrelated (column 2), correlated (column 3), and Bayesian (column 4) optimizations.}
%\label{tab:percentages}
%\end{table}

These results highlight some of the shortcomings of the frequentist methods.  The uncorrelated and correlated confidence intervals do not consistently predict the uncertainties.  Although the uncorrelated confidence intervals give a more true representation of the uncertainties at the 1$\sigma$ level -- 68\% -- for nucleon scattering on $^{40}$Ca, as shown here, in \cite{King2019}, we found that the uncorrelated frequentist optimization consistently underpredicted the uncertainties on the elastic scattering cross sections.  Thus, this test should be performed for each reaction of interest.  The model covariance, as included in $\chi^2_C$, is somewhat arbitrarily defined, pointing to possible limitations of the formulation, and we ultimately do not advocate for including model correlations in this manner.  In addition, these types of frequentists methods assume that the uncertainties can be described as a Gaussian distribution in parameter and observable space, which is not generally the case.  Particularly, this is not true for our reaction model where the dependence of the observable (differential cross section) on the parameters (optical potential) is strongly non-linear.
%First, we see that the the uncorrelated confidence intervals tend to under-predict the uncertainties , leading to an unreliable representation of the uncertainty of the optical potential in the cross section.    
%Finally, the statistical interpretation of frequentist methods needs that every possible outcome can be known, instead of being able to give the probability of a given event occurring independent of all other events - Amy i am not sure this is relevant here. in our case we are not make predictions of unique events}.

For these reasons, we have also implemented a Bayesian optimization algorithm for fitting the optical potential.  The Bayesian results (\emph{B}) are compared to the frequentist uncorrelated and correlated results in Figs. \ref{fig:UCCB} and \ref{fig:percentages}: green lines or green triangles. For the Bayesian calculations, the error $\varepsilon$ is defined again as $\varepsilon = \Delta \sigma/\overline{\sigma}$, where $\Delta \sigma$ is still the width of the 95\% confidence interval but now $\overline{\sigma}$ is the mean value of the cross sections within the 95\% confidence interval.  

In all cases, the confidence intervals and related uncertainties obtained with the Bayesian approach are larger than their frequentist counterparts (green compared to blue).  Particularly at the lower uncertainty values, the uncorrelated frequentist optimization more accurately predicts the uncertainties, but if we want to look at higher confidence values -- such as 2$\sigma$ or 95\% confidence -- the truthfulness of the uncorrelated falls off, and the Bayesian optimization becomes more reliable.  In this work and our previous studies, we look at 95\% confidence intervals, at which point the frequentist methods tend to underpredict the uncertainties and the Bayesian optimization is more truthful.   The discrepancies between the optimization methods are also seen in the propagation of the optical model uncertainties to the single-nucleon transfer cross sections, as will be discussed in Sec. \ref{sec:fewbody}.  These characteristics are not unique to nucleon scattering on $^{40}$Ca, and other targets are discussed in \cite{King2019}.
%Although we do not favor large uncertainties, we do want to report uncertainties accurately, and as shown in Fig. \ref{fig:percentages}, the Bayesian confidence intervals are consistently more accurate than either of the frequentist fits for the highest confidence intervals (which are those that interest us the most).  For $^{40}$Ca(n,n) and $^{40}$Ca(d,d), Fig. \ref{fig:percentages} (a) and (d), we see that the uncorrelated frequentist intervals follow the trend of the solid line more closely than the Bayesian optimization; however, when the frequentist trends are off, as in Fig. \ref{fig:percentages} (c) starting with the 60\% intervals, the uncertainty in the frequentist approach is significantly under-predicted.  

%%%%%%%%%%%%%%%%%%%%%%%%%%%%%

\subsection{Correlations in parameter space}
\label{sec:corr}

It is worth also discussing the differences between the minima and correlations in parameter space among the three methods.  In Table \ref{tab:parameters}, we show the best-fit parameters for the uncorrelated and correlated frequentist fits along with the average parameter values from the accepted Bayesian samples.   We note that, especially in the imaginary volume term, many of the parameters have to be fixed in order to prevent the minimum from going into unphysical regions of the parameter space when the frequentist optimization is used.  These parameters are shown in italics in Table \ref{tab:parameters}.  However, the real volume term is fairly similar between the three optimization schemes, meaning that they all lead to similar minima, the biggest differences being the geometries of the potentials. Moreover, particularly in the frequentist optimization, the parameters tend to be highly correlated, leading to similar elastic-scattering cross sections, even when the geometry is not the same.  We can also constrain the Bayesian optimization to have the same parameters fixed (at the same values) as the uncorrelated frequentist optimization, and when this is done, we find essentially the same minima between the two routines.  We then conclude that the prior distribution -- especially a very wide Gaussian prior, as we use here -- does not strongly affect the minima that are found in the Bayesian optimization and does not necessarily keep the minimum closer to the starting parameter set.  The Bayesian optimization procedure is data-driven (or driven by the likelihood) rather than being driven by the prior distribution, and the Bayesian procedure allows the parameters to remain in the physical region of the parameter space, while adding a very minimal constraint with the prior distribution.

In the bottom half of Table \ref{tab:parameters}, we list the widths of the parameter distributions, either from the parameter covariance matrix for the frequentist calculations or the standard deviation of the set of accepted parameters for the Bayesian calculations.  In most cases, the Bayesian parameters widths are larger than both the uncorrelated and correlated frequentist widths.  The largest difference is for the correlated frequentist optimization, where sometimes the imaginary surface parameters are wider than the parameters found in the Bayesian optimization.  Noticeably, the optimization of $^{40}$Ca(p,p) at 14.5 MeV is the only case where all of the Bayesian parameter widths are larger than the correlated frequentist parameters, and this is the only case where the Bayesian confidence intervals are larger than the correlated confidence intervals across the full angular range.

There is a high level of degeneracy in the parameter space, both continuous and discrete, due to the fact that different sets of optical parameters can provide the same elastic scattering distribution. One way to address this degeneracy is to  introduce new parameters resulting from a combination of optical potential parameters that would remove the degeneracy. This is not our approach. In fact, because we choose to use wide priors, we explore a wide region of parameter space, with all its degeneracies, and allow for multimodal posteriors. Our uncertainty intervals thus have no biases toward one or another minimum. Nevertheless, by construction the priors do not favor unphysical value for the parameters, avoiding issues with negative values for radii and diffuseness. 

\begin{landscape}
\begin{table}
\centering
\begin{tabular}{clccccccccc}
\textbf{Optimization} & \textbf{Reaction} & \textbf{V ($\mu$)} & \textbf{r ($\mu$)} & \textbf{a ($\mu$)} & \textbf{W$_\mathrm{s}$ ($\mu$)} & \textbf{r$_\mathrm{s}$ ($\mu$)} & \textbf{a$_\mathrm{s}$ ($\mu$)} & \textbf{W$_\mathrm{v}$ ($\mu$)} & \textbf{r$_\mathrm{W}$ ($\mu$)} & \textbf{a$_\mathrm{W}$ ($\mu$)} \\ \hline 
\textbf{UC} & $^{40}$Ca(n,n) (13.9) &45.629 &1.293 &0.500 &3.959 &1.129  &\it{0.890} &\it{1.289} &\it{1.094} &\it{0.301} \\
\textbf{C} & $^{40}$Ca(n,n) & 43.284 &1.316 &0.532 & 4.303 &1.111 & 0.789 &\it{1.301} &\it{1.332} &\it{0.652} \\
\textbf{B} & $^{40}$Ca(n,n) & 44.246 & 1.317 & 0.550 & 7.268 & 1.231 & 0.484 & 1.276 & 1.095 &0.587 \\ \hline
\textbf{UC} & $^{40}$Ca(p,p) (14.5) &53.286 &1.160 &0.590 &2.145 &1.279 &\it{0.892} &\it{1.189} &\it{1.070} &\it{0.757}  \\
\textbf{C} & $^{40}$Ca(p,p) & 52.467 &1.267 &0.488 &2.008 &\it{1.045} &\it{0.7849} &\it{2.088} &\it{1.118} &\it{0.652} \\
\textbf{B} & $^{40}$Ca(p,p) & 51.472 & 1.207 & 0.639 & 6.471 & 1.312 & 0.371 & 0.425& 1.335 & 0.530 \\ \hline
\textbf{UC} & $^{40}$Ca(p,p) (26.3) & 55.776 &1.067 &0.881 &4.206 &\it{1.034} &\it{0.721} &\it{4.421} &\it{1.3121} &\it{0.51} \\
\textbf{C} & $^{40}$Ca(p,p) & 44.676 &1.258 &0.7077 & 6.7437 &1.210 & 0.403 & \it{2.588} &\it{1.362} &\it{0.7556} \\
\textbf{B} & $^{40}$Ca(p,p) & 56.538 & 1.025 & 0.865 & 3.027 & 1.591 & 0.602 & 3.408 & 1.390 & 0.483 \\ \hline
\textbf{UC} & $^{40}$Ca(d,d) (30.0) & 97.936 &1.065 &0.805 &9.045 &1.516 &0.630 &\it{2.444} &\it{1.479} &\it{0.311} \\
\textbf{C} & $^{40}$Ca(d,d) & 100.49 &1.077 &0.774 &\it{6.624} &1.413 &0.853 &\it{2.010} &\it{1.600} &\it{0.334} \\
\textbf{B} & $^{40}$Ca(d,d) & 97.026&1.080&0.782&9.697&1.486&0.676&2.617&1.045&0.555 \\  \hline 
 & & \textbf{V ($\sigma$)} & \textbf{r ($\sigma$)} & \textbf{a ($\sigma$)} & \textbf{W$_\mathrm{s}$ ($\sigma$)} & \textbf{r$_\mathrm{s}$ ($\sigma$)} & \textbf{a$_\mathrm{s}$ ($\sigma$)} & \textbf{W$_\mathrm{v}$ ($\sigma$)} & \textbf{r$_\mathrm{W}$ ($\sigma$)} & \textbf{a$_\mathrm{W}$ ($\sigma$)} \\ \hline 
\textbf{UC} & $^{40}$Ca(n,n) (13.9) &2.426 &0.045 &0.048 &0.722 &0.185 & --- & --- & --- & --- \\
\textbf{C} & $^{40}$Ca(n,n) & 2.022 &0.029 &0.076 &1.210 &0.203 &0.237 & --- & --- & --- \\
\textbf{B} & $^{40}$Ca(n,n) & 3.277 & 0.057 & 0.057 & 0.604 & 0.086 & 0.038 & 0.150 & 0.106 & 0.069 \\ \hline
\textbf{UC} & $^{40}$Ca(p,p) (14.5) &1.475 &0.016 &0.024 &0.309 &0.170 & --- & --- & --- & --- \\
\textbf{C} & $^{40}$Ca(p,p) & 1.889 &0.020 &0.017 &0.202 & --- & --- & --- & --- &  ---- \\
\textbf{B} & $^{40}$Ca(p,p) & 2.915 & 0.047 & 0.060 & 0.648 & 0.114 & 0.053 & 0.046 & 0.154 & 0.059 \\ \hline
\textbf{UC} & $^{40}$Ca(p,p) (26.3) & 2.245 &0.027 &0.023 &0.210 & --- & --- & --- & --- & --- \\
\textbf{C} & $^{40}$Ca(p,p) &1.062 & 0.027 &0.031 &1.243 &0.035 &0.050 & --- & --- & --- \\
\textbf{B} & $^{40}$Ca(p,p) & 3.622 & 0.044 & 0.050 & 0.446 & 0.096 & 0.058 & 0.285 & 0.098 & 0.058 \\ \hline
\textbf{UC} & $^{40}$Ca(d,d) (30.0) & 5.361 &0.043 &0.023 &0.511 &0.015 &0.026 & --- & --- & ----  \\
\textbf{C} & $^{40}$Ca(d,d) & 5.743 &0.127 &0.095 & --- &0.155 &0.148 & --- & --- & --- \\
\textbf{B} & $^{40}$Ca(d,d) & 7.269 & 0.056 & 0.038 & 0.950 & 0.042 & 0.054 & 0.299 & 0.144 & 0.051 \\
\end{tabular}
\caption{Optimized parameter means, $\mu$, and and posterior widths, $\sigma$, for each of the three fitting techniques (first column) and each reaction studied (incident energies, in MeV, listed in parentheses in the second column).  The real volume, imaginary surface, and imaginary volume depths are listed, in MeV, in the third, sixth, and ninth columns, respectively.  The corresponding radii (diffusenesses) are listed in fm in the fourth (fifth), seventh (eighth), and tenth (eleventh) columns.}
\label{tab:parameters}
\end{table}
\end{landscape}

\begin{figure}
\centering
\begin{tabular}{cc}
\pdfimageresolution=100\includegraphics[width=0.45\textwidth]{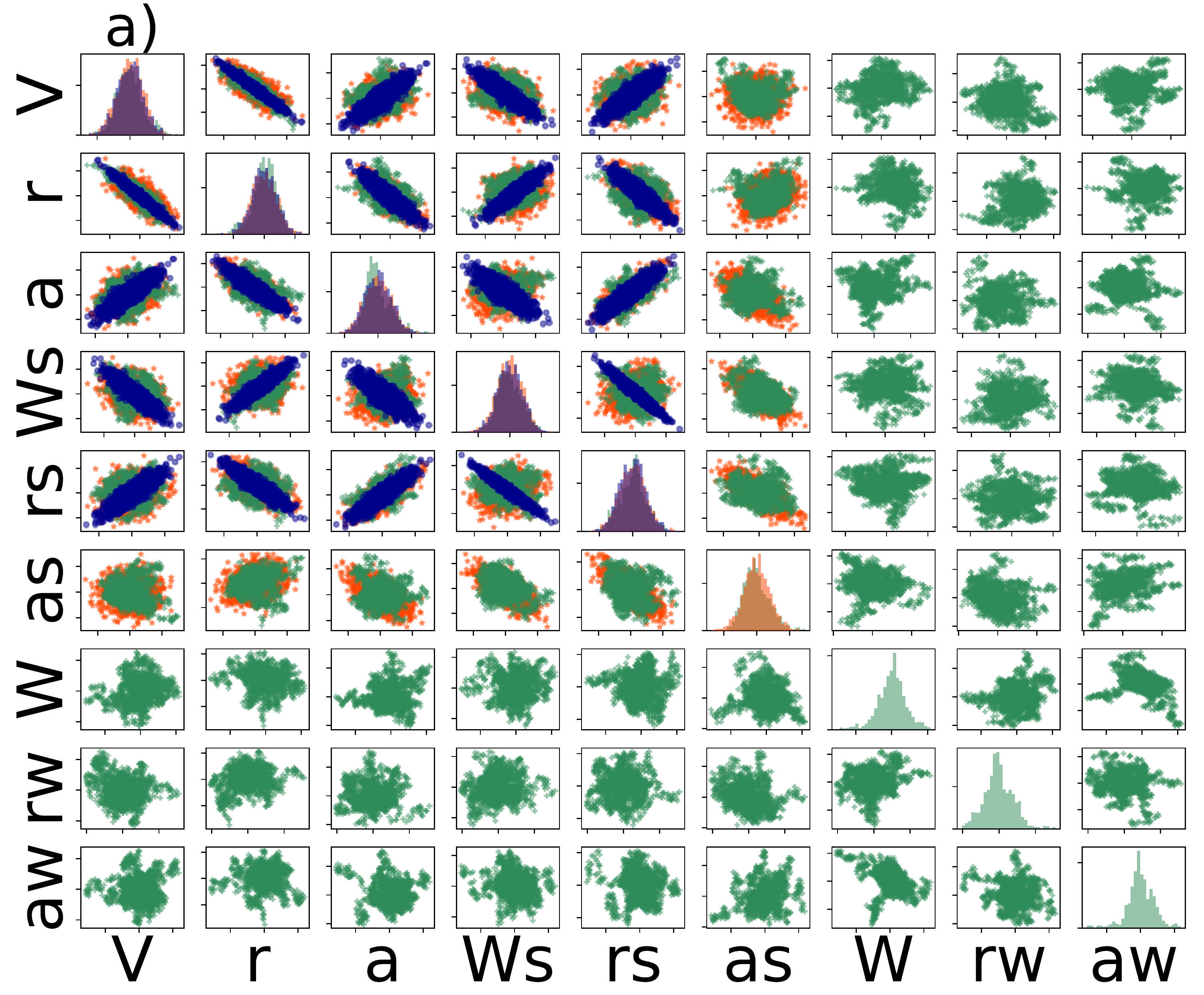} &
\pdfimageresolution=100\includegraphics[width=0.45\textwidth]{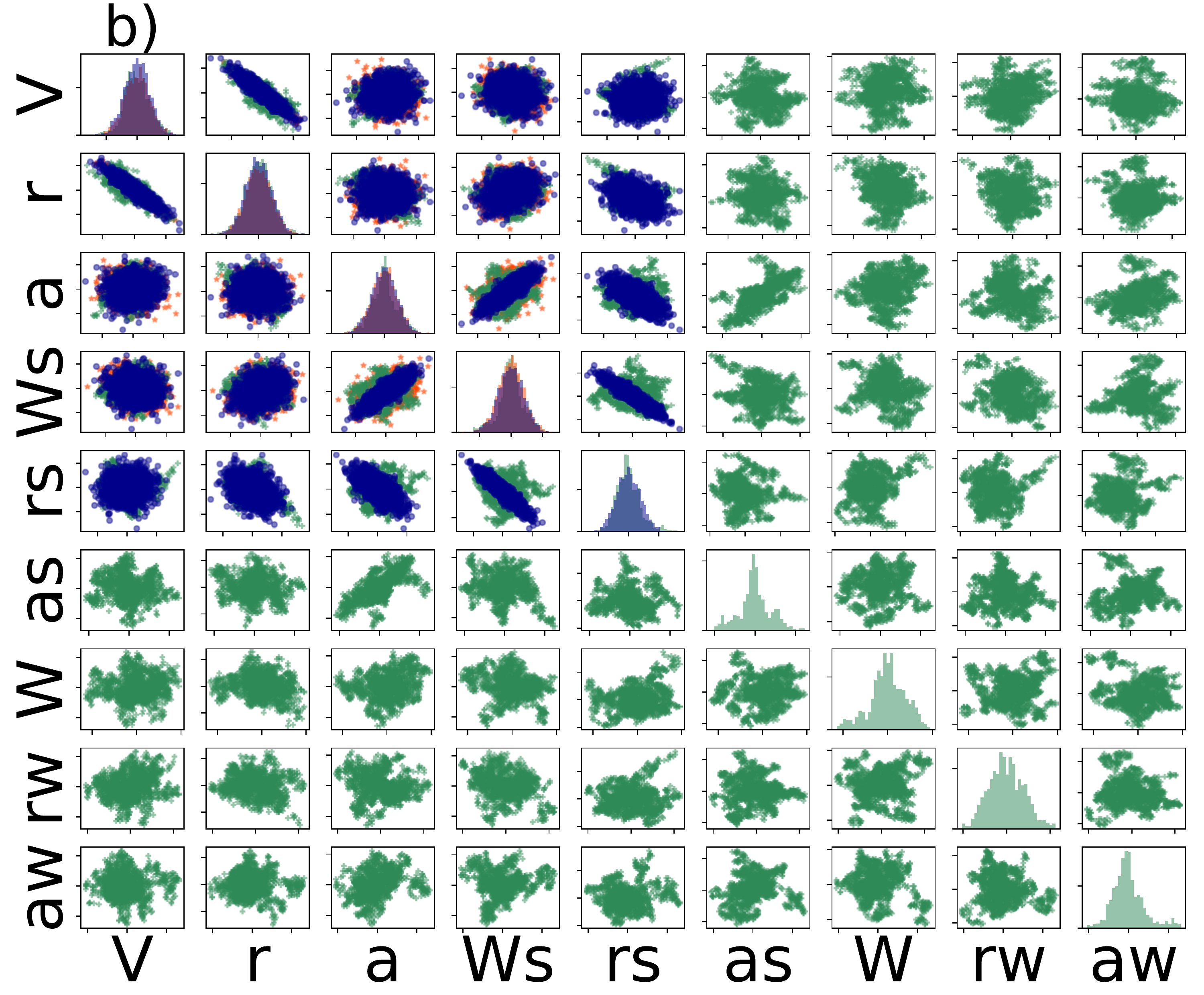} \\
\pdfimageresolution=100\includegraphics[width=0.45\textwidth]{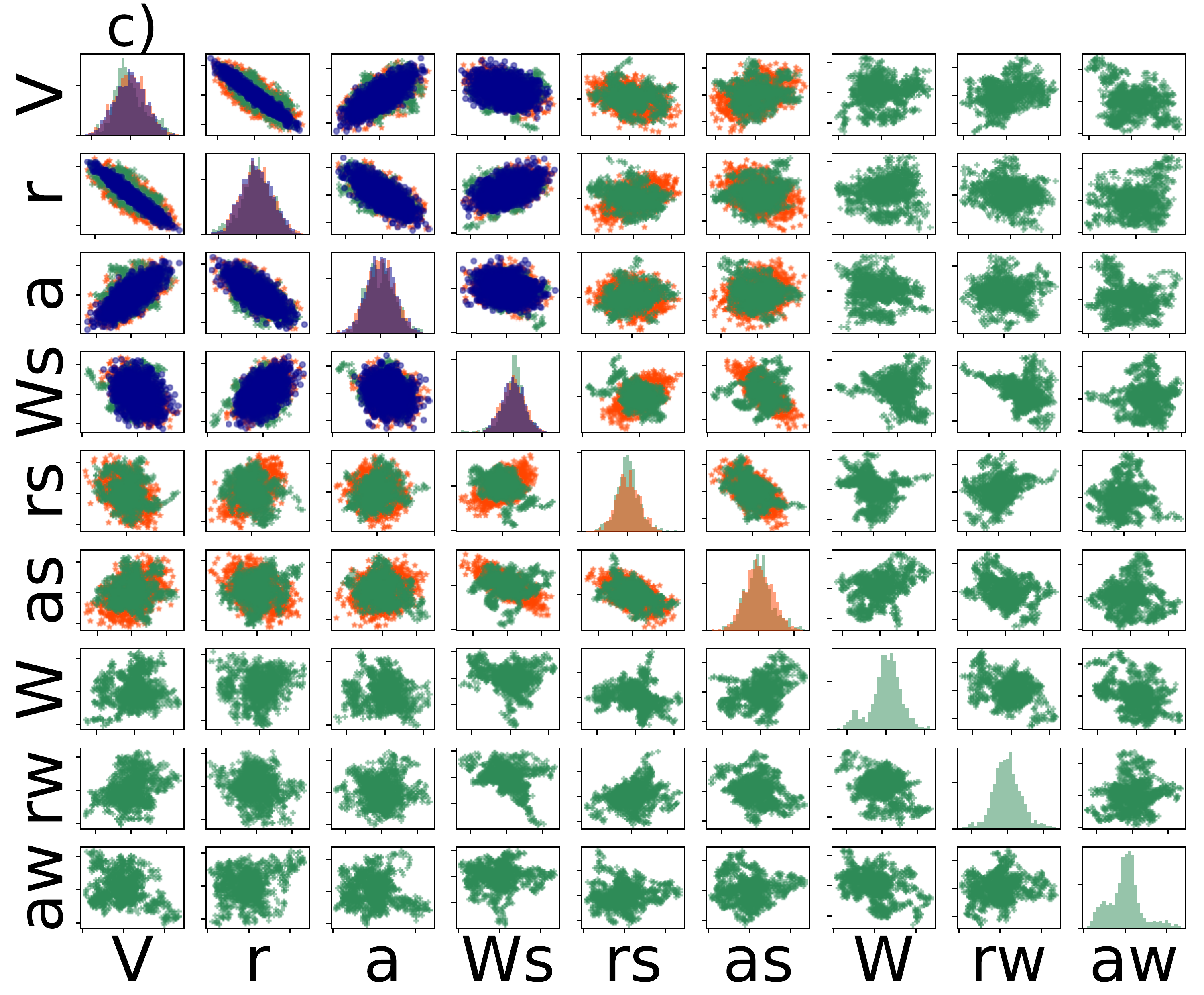} &
\pdfimageresolution=100\includegraphics[width=0.45\textwidth]{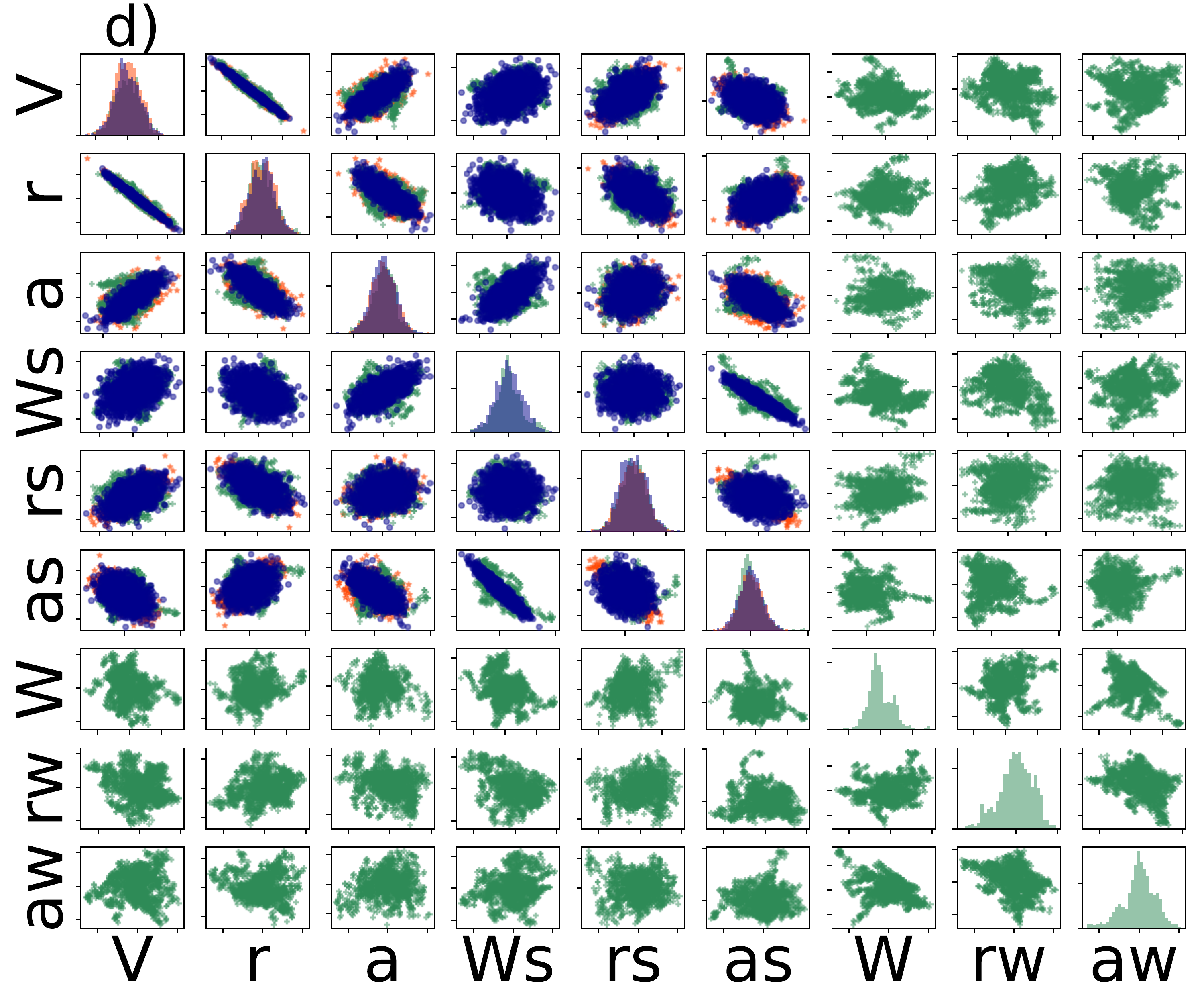} \\
\end{tabular}
\caption{Correlations between the optical model parameters for (a) $^{40}$Ca(n,n) at 13.9 MeV, (b) $^{40}$Ca(p,p) at 14.5 MeV, (c) $^{40}$Ca(p,p) at 26.3 MeV, and (d) $^{40}$Ca(d,d) at 30.0 MeV for the uncorrelated, correlated, and Bayesian optimizations in blue, red, and green respectively.  Histograms on the diagonal and parameters in the off-diagonal scatter plots have been normalized to mean zero and standard deviation one to emphasize the differences in the correlations.}
\label{fig:correlations}
\end{figure}

In addition, the correlations between the parameters from the four elastic-scattering optimizations are shown in Fig. \ref{fig:correlations}.  To focus on the differences in the correlations and not the parameter values, the correlations have been normalized such that each mean is zero and the width of the distribution is one.  Circular distributions indicate uncorrelated parameters while the more oval distribution indicate more correlated (or anti-correlated) parameter pairs.  Historically, the optical model parameters had been found to be extremely correlated \cite{ThompsonNunes}, and these correlations can be seen in the blue and red distributions of the uncorrelated and correlated parameters.  On the other hand, the Bayesian optimization shows very few correlations, except between the depth and radius of the real volume potential.  This lack of correlation was also seen in \cite{King2019} and indicates that strong correlations may have been induced by the $\chi^2$ minimization procedure.  

Now that we have seen the uncertainties coming from the optical potential with the constraints coming from fitting the differential cross sections, we find that these uncertainties are so large as to be almost unusable.  The uncertainties that we have found in all studied cases have varied anywhere from 20\% to over 100\% - and are not any smaller when they are propagated to the single-nucleon transfer cross sections.  Therefore, we have also been exploring ways to reduce the uncertainties in the optical potential.  In the next two subsections, we discuss further experimental constraints to shrink the uncertainties on the optical potential - and the resulting uncertainties on the elastic scattering and transfer cross sections.

\subsection{Other experimental constraints on the optical potential}
\label{sec:energies}

Next we explore experimental conditions for elastic-scattering measurements, with the intent of reducing the uncertainties. Here we  include tests on the angular range of the data fitted and the effects of including a second set of elastic  angular distributions, obtained at a nearby energy.  

The angular range of the data included in the optimization procedure in \cite{CatacoraRios2019}, tested whether fitting only angles forward of 100$^\circ$ or fitting a reduced set of angular data (where every other angular data point was removed) produced significant differences in the width of the uncertainty interval.  Because of the correlations between the differential cross section at various angles, due to the partial wave decomposition, constrains at one angle are propagated to other angles.  Consistent with that, we find that we do not gain more information with a denser angular grid, and constraining backwards angles expectedly only affects the backward angle uncertainty.  

\begin{figure}
\centering
\begin{tabular}{cc}
\includegraphics[width=0.5\textwidth]{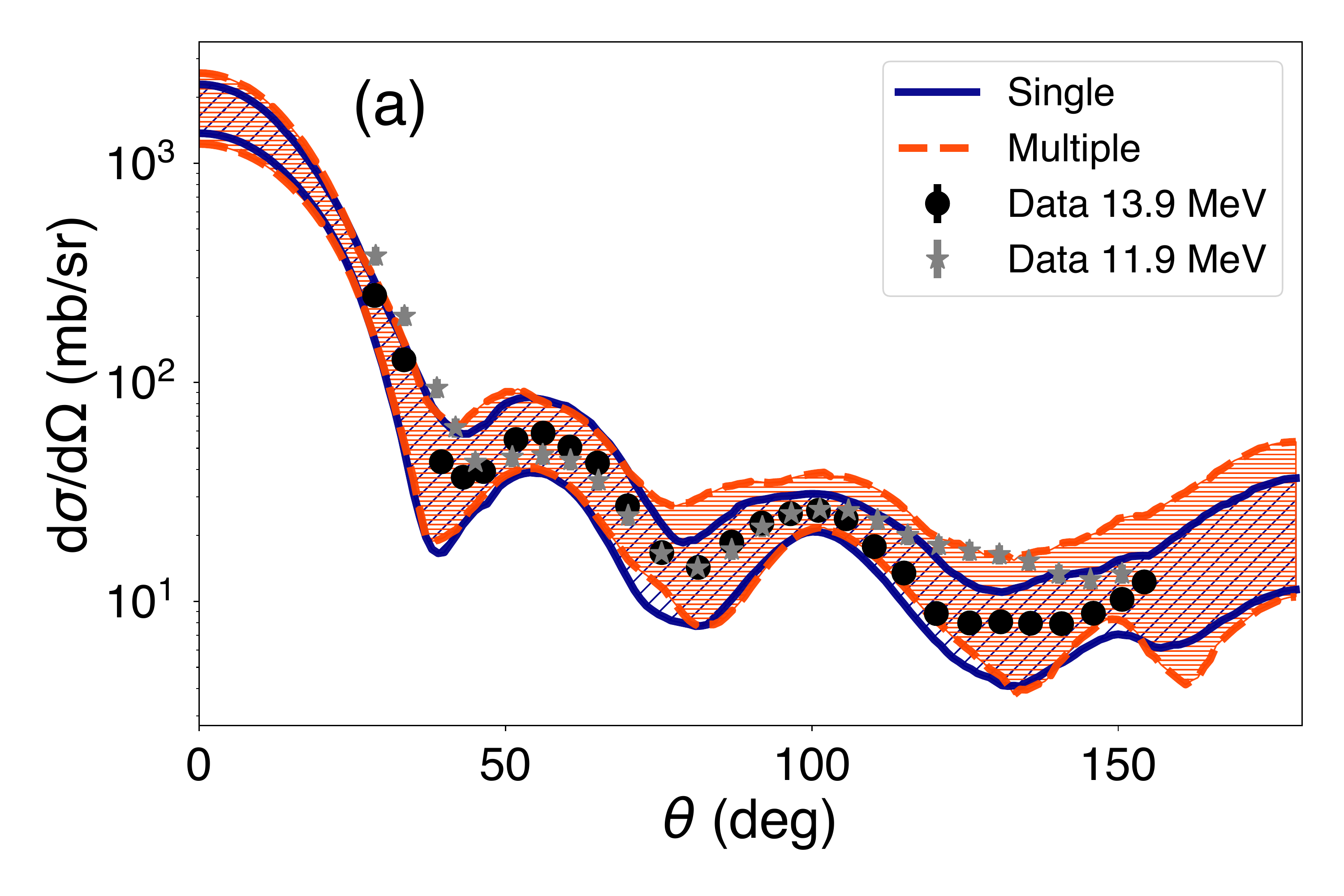} & \includegraphics[width=0.5\textwidth]{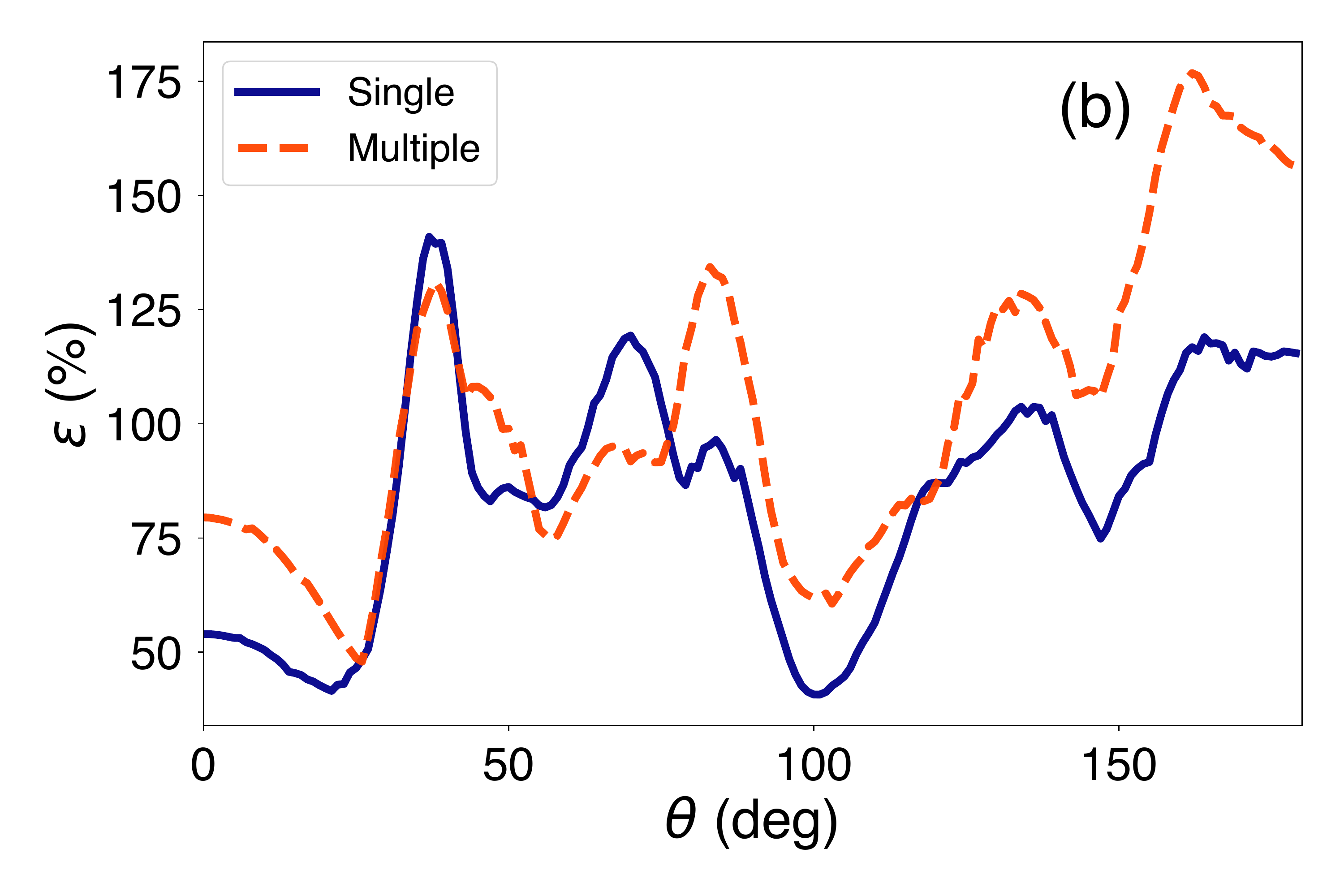} \\
\includegraphics[width=0.5\textwidth]{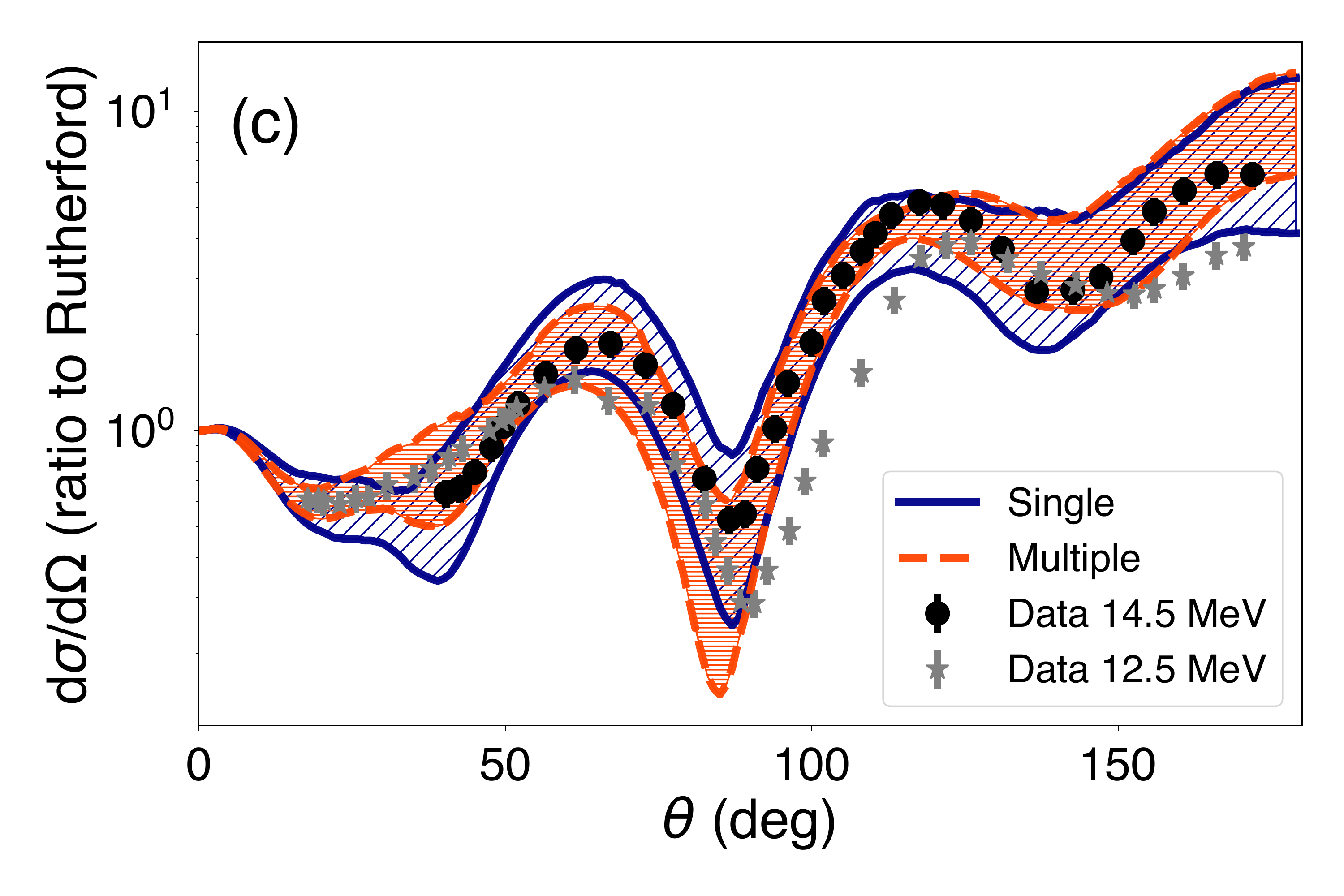} & \includegraphics[width=0.5\textwidth]{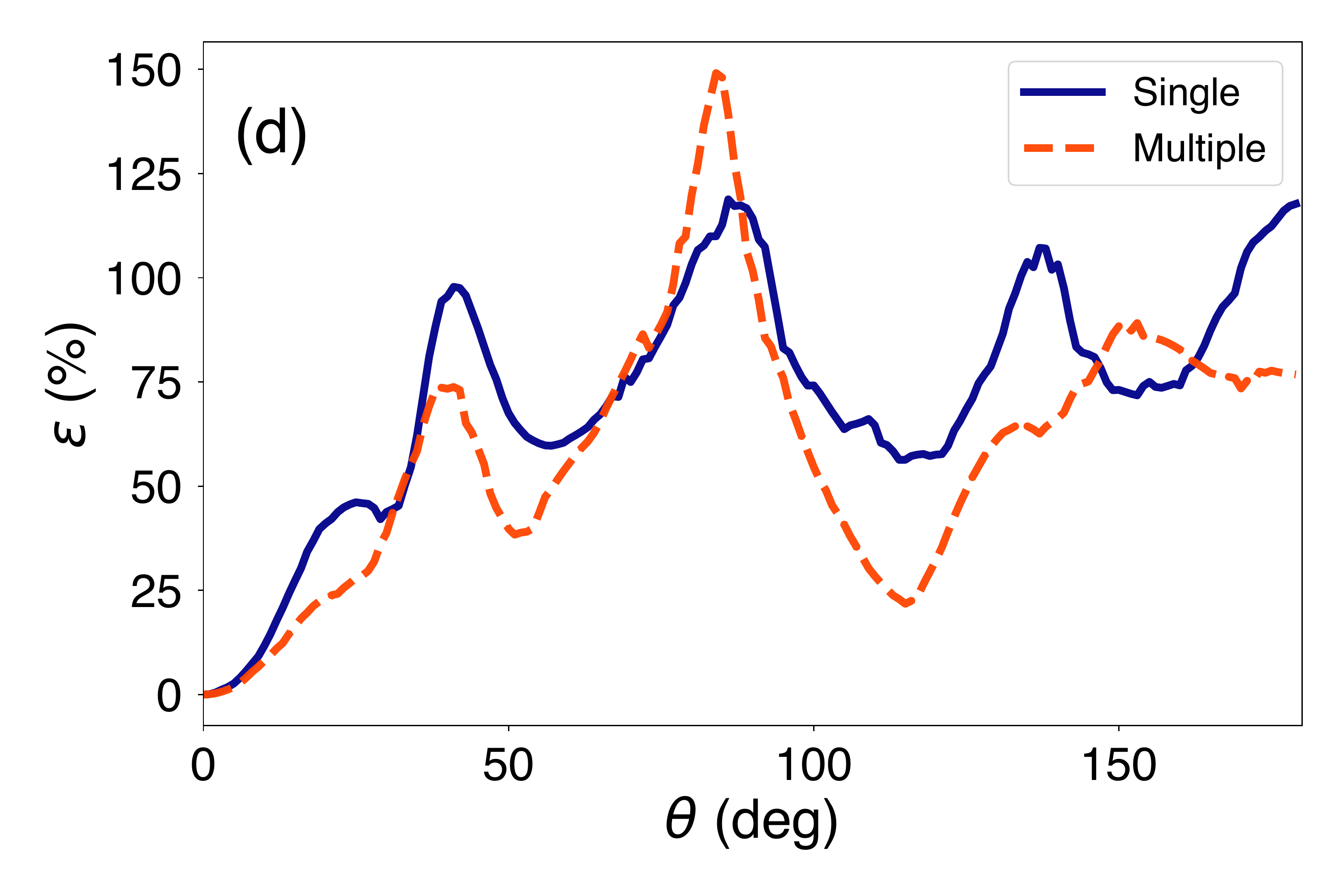} \\
\end{tabular}
\caption{Comparison between the Bayesian optimization of a single elastic scattering data set (\emph{single}, blue) and the simultaneous optimization of two nearby energies (\emph{multiple}, red) for (a) $^{40}$Ca(n,n) at 13.9 MeV (second set at 11.9 MeV), and (c) $^{40}$Ca(p,p) at 14.5 MeV (second set at 12.5 MeV).  In (b) and (d) are the corresponding percentage uncertainties, $\varepsilon$.}
\label{fig:energies}
\end{figure}

Instead, we turn to including a second set of elastic-scattering data at a nearby energy.  There are two ways to include this second set:  1) sequentially, where the Bayesian optimization is run using the first set of data and the parameter posterior distribution is used as the prior distribution to optimize over the second set of data, or 2) simultaneously, where the two data sets are both fed into the optimization routine at the same time.  When the two data sets are included sequentially, we find a very small improvement in the uncertainties (decrease in the size of the confidence intervals), just as in \cite{CatacoraRios2019}.  However, while \cite{CatacoraRios2019} shows a  large improvement when the two data sets are included simultaneously, in the cases studied here we find only a modest improvement.  This is illustrated in Fig. \ref{fig:energies}: a simultaneous fit (denoted \emph{multiple}, red) is compared to the fitting of only one data set (\emph{single}, blue) in for $^{40}$Ca(n,n) at 13.9 MeV (a) and $^{40}$Ca(p,p) at 14.5 MeV (c).  In panels (b) and (d), we show the percent uncertainty of the confidence intervals, $\varepsilon$.  %\IR{Perhaps, in this case, we also want to show the sequential case, since there is not much difference here.}

As is evident from the percent uncertainty in panels (b) and (d) for the neutron scattering and proton scattering respectively, no significant decrease in the uncertainty is found when the second set of experimental data is included.  In particular, for the neutron scattering case, we actually see an increase of the uncertainties at backwards angles.  This increase is due to the discrepancies  between the two data sets backwards of $100^\circ$ (shown as the black and grey symbols).  In both cases, we used data that was measured by the same group with the same experimental set-up, to remove as many sources of systematic uncertainty as possible.  The differences between the results here and the results in our previous work \cite{CatacoraRios2019} are mainly due to the use of mock data in the previous work and real experimental data here.  We discuss this further in Section \ref{sec:dataDifferences}.

\subsection{Including a variety of reaction observables}

Within the Bayesian framework, we can study how various additional experimental constraints can impact - and hopefully reduce - the uncertainties in the optical potential.  The first experimental constraint that we explore is adding other reaction data to the fitting procedure. We will start first considering vector analyzing powers, $Re(iT_{11})$, and then add total or reaction cross sections, $\sigma_\mathrm{tot}$ or $\sigma_\mathrm{R}$.  %We take 10\% error on all data points except for when the values of polarization $Re(iT_{11})<0.05 \times \mathrm{Max}(Re(iT_{11}))$. For those intervals we assigned a flat error  $\varepsilon =0.05 \times \mathrm{Max}(Re(iT_{11}))$.  
The combined $\chi^2$ applies equal weights to the different sets of data.  Although the vector analyzing powers should be more sensitive to the spin-orbit part of the potential than the differential cross sections or the total and reaction cross sections, for consistency within this work, we still only optimize the volume and surface terms of the optical potential.

\begin{figure}
\centering
\begin{tabular}{cc}
\includegraphics[width=0.5\textwidth]{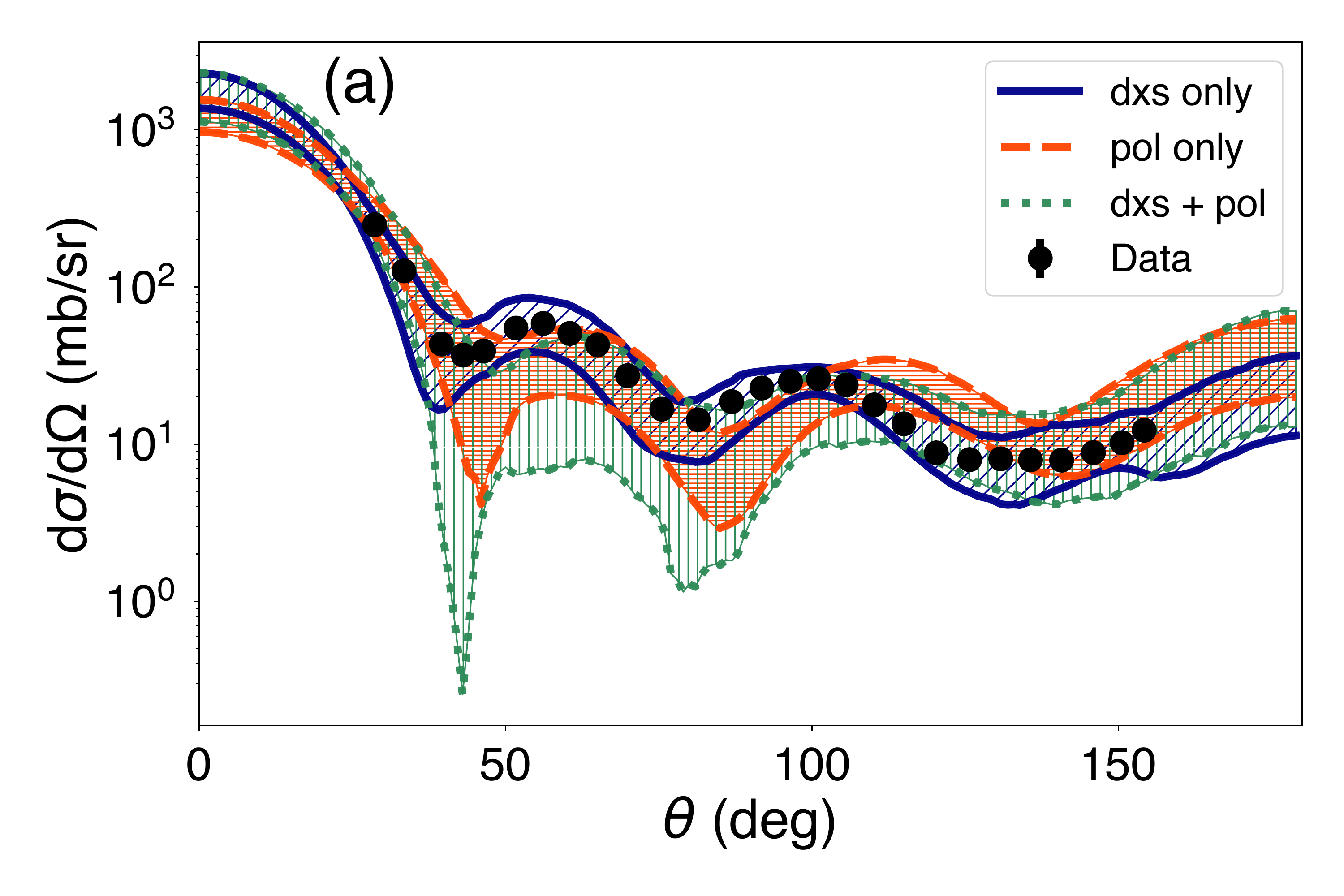} & \includegraphics[width=0.5\textwidth]{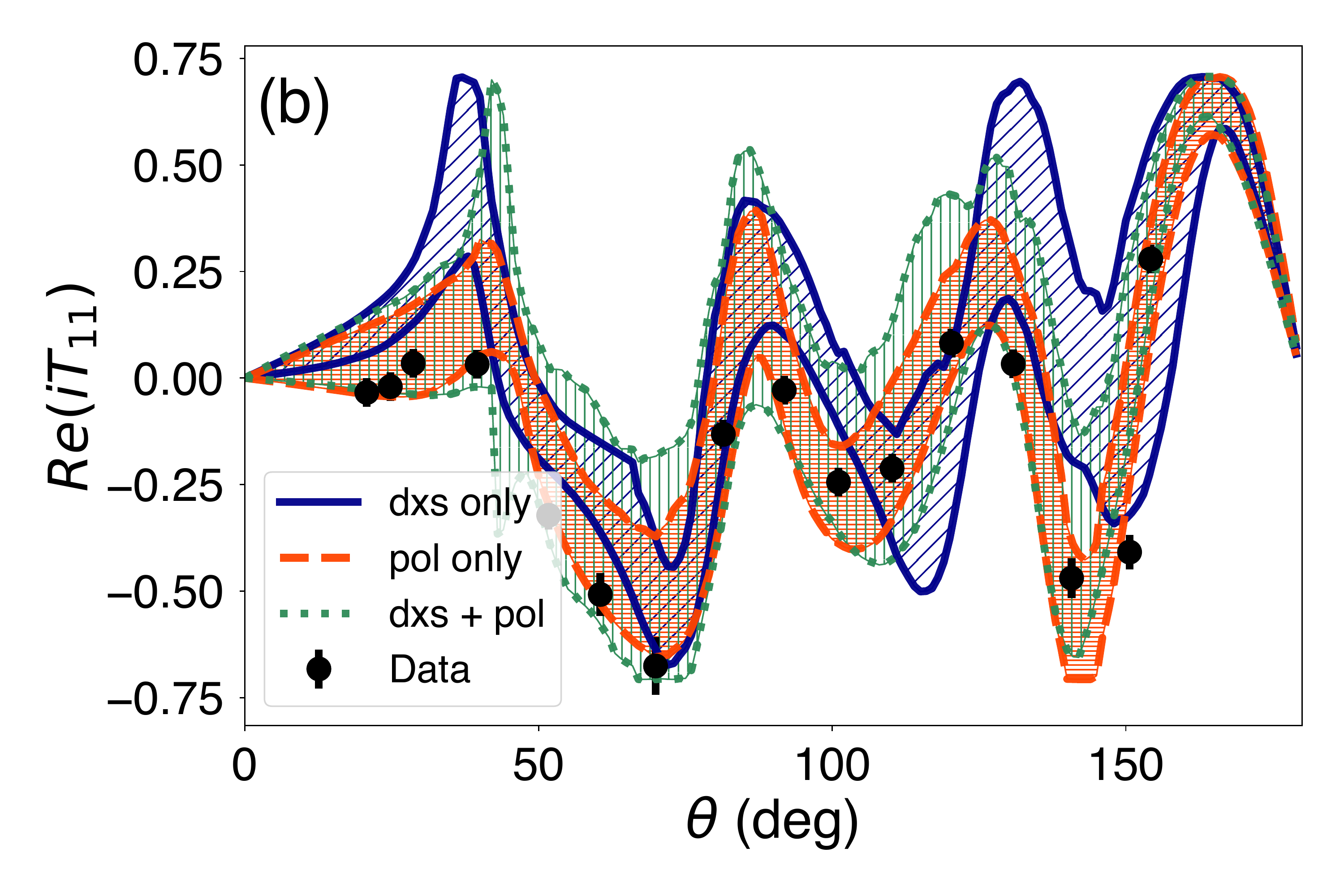} \\
\includegraphics[width=0.5\textwidth]{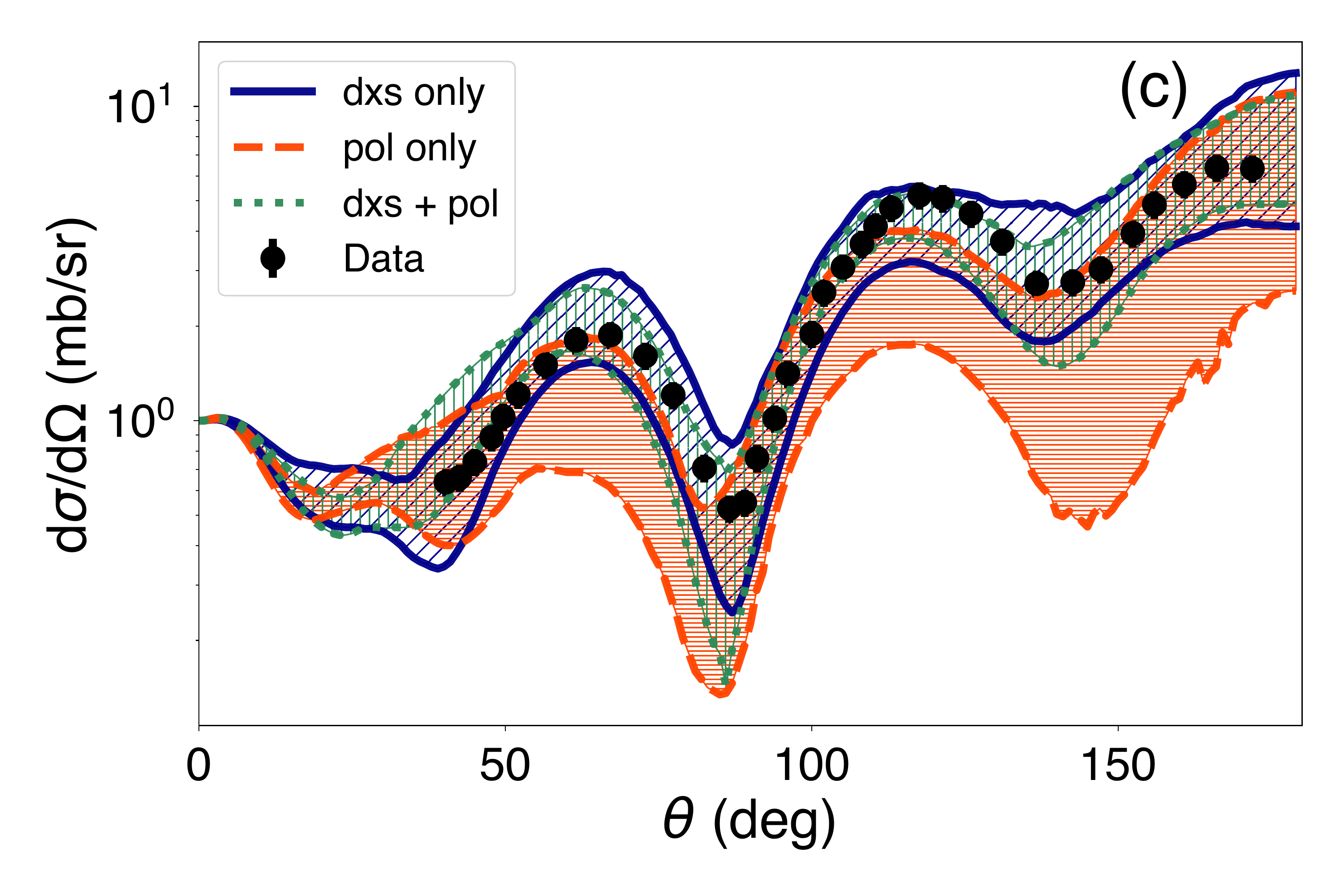} & \includegraphics[width=0.5\textwidth]{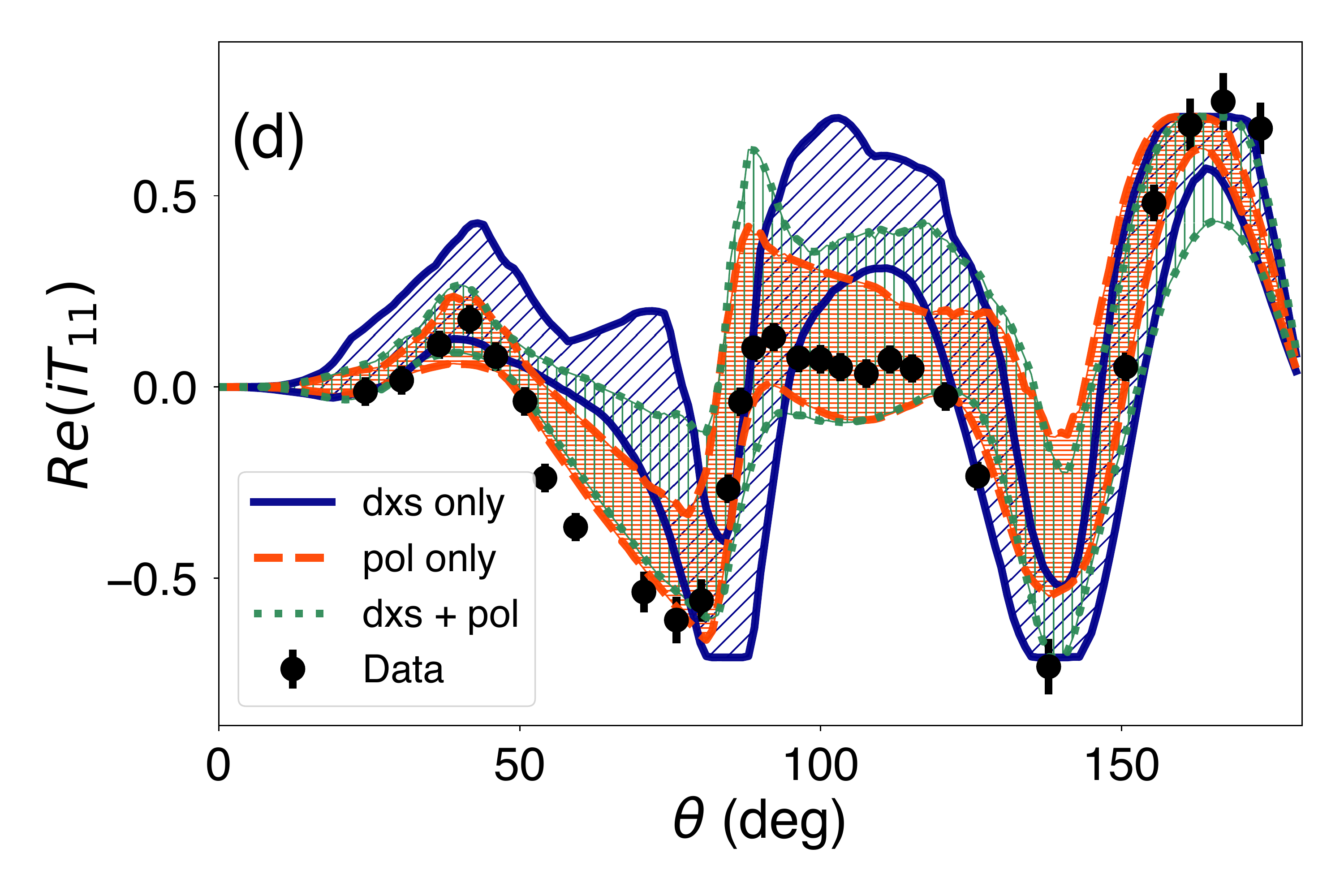} \\
\end{tabular}
\caption{Comparison of the differential cross sections (a) and (c) and polarization analyzing powers (b) and (d) when only differential cross section data was fit (blue), only polarization data was fit (red), and both differential cross section and polarization data were fit, for $^{40}$Ca(n,n) at 13.9 MeV, (a) and (b), and for $^{40}$Ca(p,p) at 14.5 MeV, (c) and (d).}
\label{fig:observables}
\end{figure}

In Fig. \ref{fig:observables}, we show the differential cross sections (left column) and analyzing powers (right column) when only the differential cross section is fit (blue), only the polarization is fit (red), and both are fit simultaneously (green) for $^{40}$Ca(n,n) at 13.9 MeV and $^{40}$Ca(p,p) at 14.5 MeV.  First, we notice that in (a) and (c) the best representation of the data is when only the differential cross section is fit; likewise, in panels (b) and (d), the best representation of the polarization data is when only the polarization data is included in the optimization.  These are the sorts of difficulties that can arise when addressing the real problem, with real data. In both the neutron and proton scattering cases, we see that the minima are shifted significantly from the differential cross section to the polarization minimum.  When both observables are included in the Bayesian optimization (green regions), the confidence intervals and posterior distributions typically fall somewhere between the minima from the elastic cross section and polarization, as for $^{40}$Ca(p,p) at 14.5 MeV in Fig. \ref{fig:observables}(c) and (d).  However, as for $^{40}$Ca(n,n) at 13.9 MeV, in panels (a) and (b), we find that the confidence intervals when both observables are included in the fitting process are similar in shape to the confidence intervals when only the analyzing power is included in the optimization.  The authors in \cite{Tornow1982}, \cite{Aoki1996}, and , \cite{Mccamis1986} all found that for $^{40}$Ca-nucleon scattering reactions, an optical potential of the form that we are investigating here (e.g. local, energy-dependent, and non-relativistic) was not sufficient to describe both their differential cross section data and polarization data simultaneously.  These joint optimizations should be studied for other targets.

We have also considered the effect of including total or reaction cross sections along with polarization.  In \cite{CatacoraRios2019}, we found that adding the reaction or total cross section into the Bayesian optimization routine did not offer an improvement on the uncertainties, when compared to the use of the full angular distribution. In that work, mock data was used and thus the total (reaction) cross section did not contain additional information to the differential cross section over all angles (by construction that total cross section was exactly the integral of the angular distribution included in the fit already). Although here we are using real data, we have chosen cases for which the angular distributions have full coverage, and therefore again, when including total (reaction) cross sections to the likelihood, the reduction on the uncertainties is minimal.

\subsection{Differences between mock data and measured data}
\label{sec:dataDifferences}

%\IR{FN: we need to discuss this in our meeting!!! there is no reason to not do additional calculations to answer the questions here posed}
 
\begin{figure}
\centering
\begin{tabular}{cc}
\includegraphics[width=0.5\textwidth]{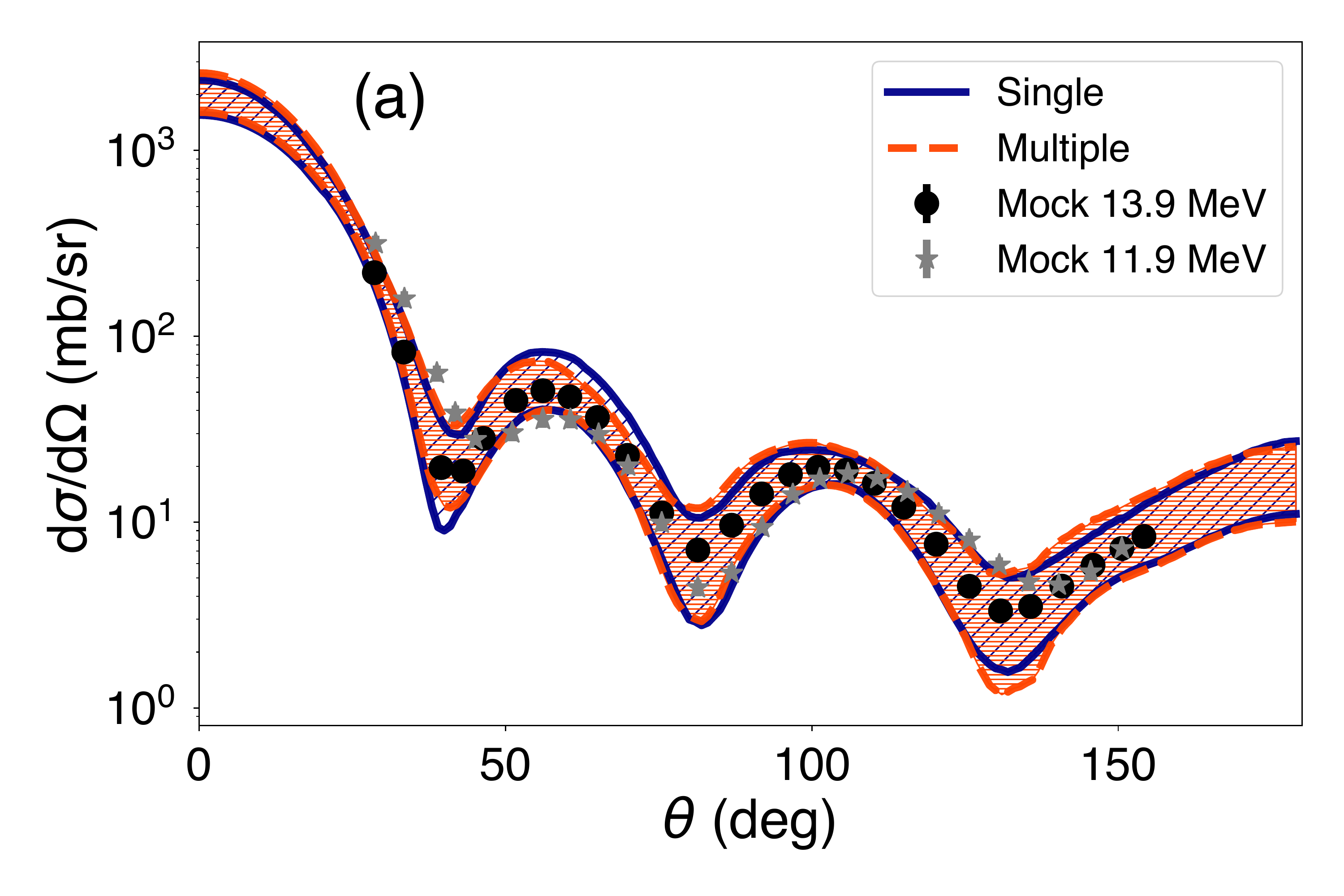} & \includegraphics[width=0.5\textwidth]{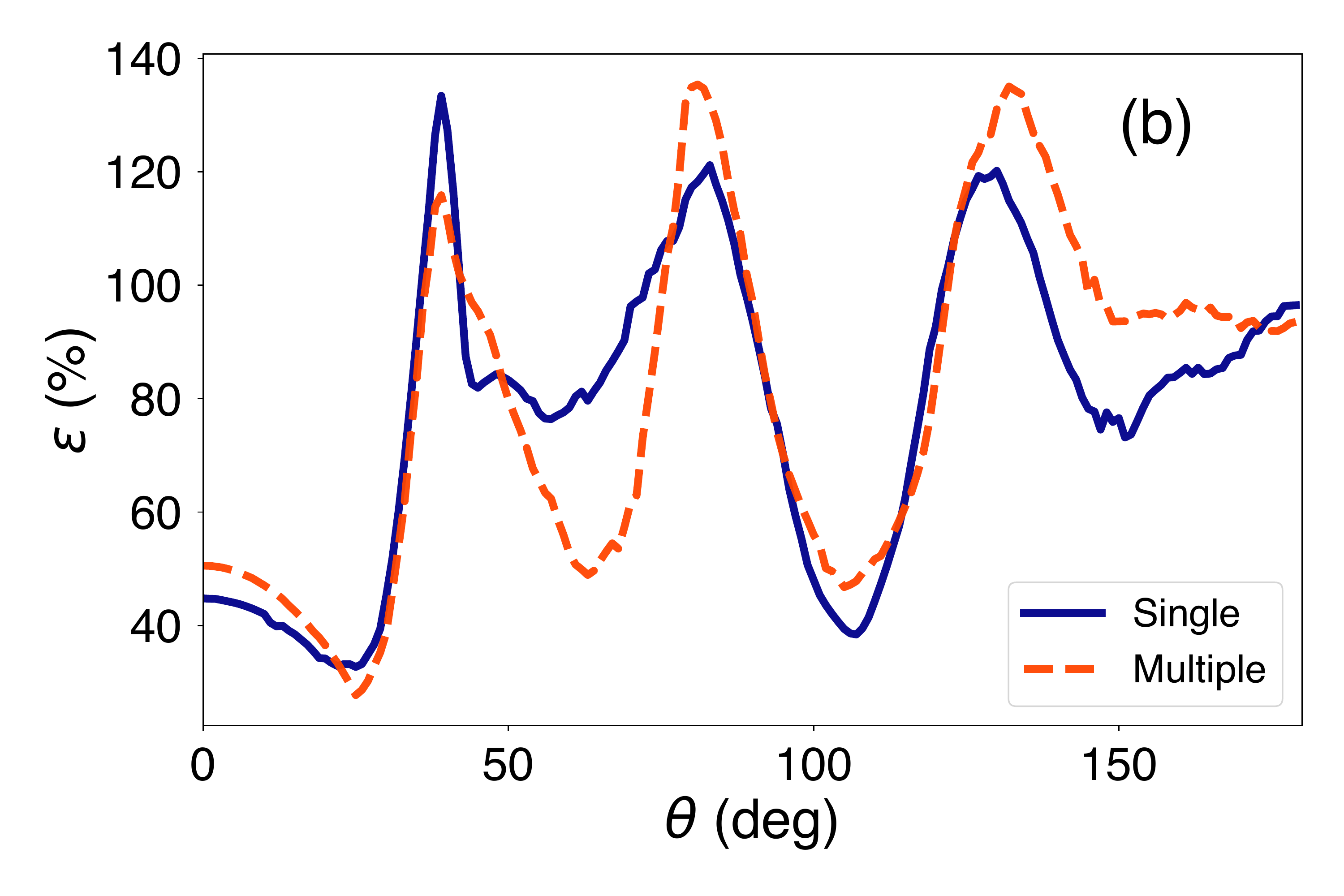} \\
\includegraphics[width=0.5\textwidth]{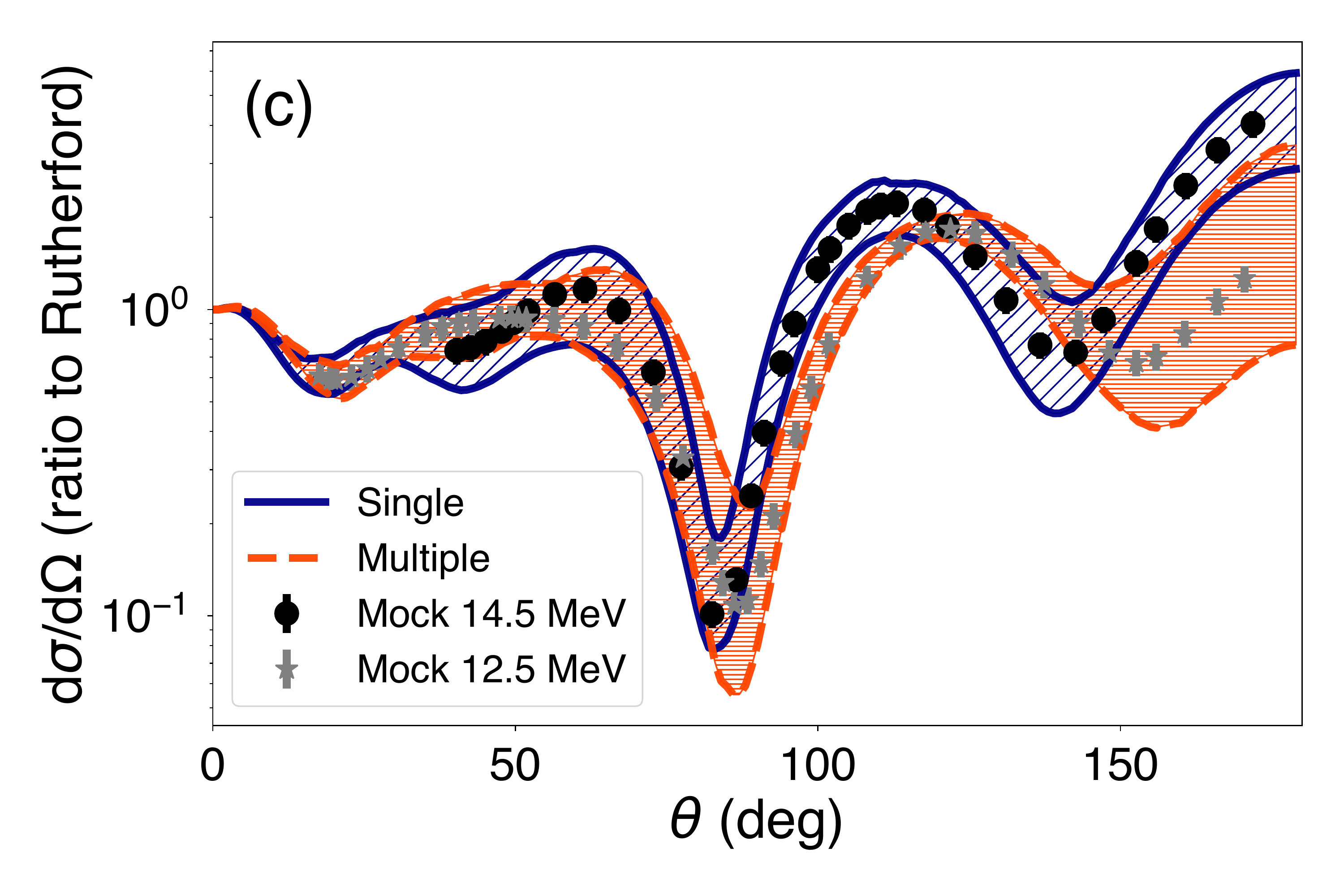} & \includegraphics[width=0.5\textwidth]{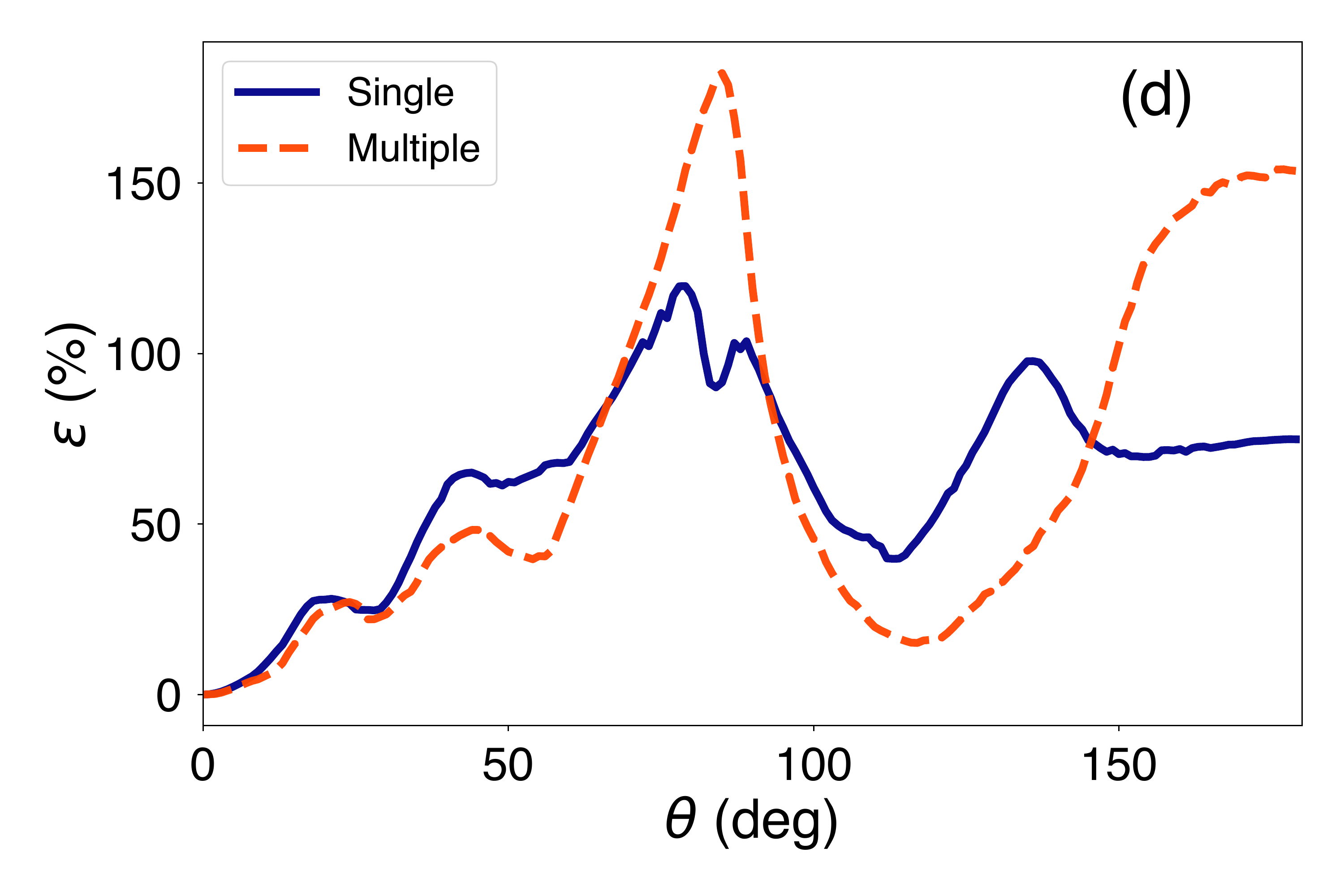} \\
\end{tabular}
\caption{Comparison between the Bayesian optimization of a single elastic scattering data set (\emph{single}, blue) and the simultaneous optimization of two nearby energies (\emph{multiple}, red) for (a) $^{40}$Ca(n,n) mock data at 13.9 MeV (second set at 11.9 MeV), and (c) $^{40}$Ca(p,p) mock data at 14.5 MeV (second set at 12.5 MeV).  In (c) and (d) are the corresponding percentage uncertainties, $\varepsilon$.}
\label{fig:energiesMock}
\end{figure}

Some of the results presented in this section are significantly different than those obtained in  \cite{CatacoraRios2019}, particularly when multiple energies were included in the optimization.  We had previously found that fitting data at two nearby energies reduced the uncertainty on the differential cross section by as much as - and sometimes more than - 50\%.  Those results were calculated using mock data generated from the Koning-Delaroche optical potential \cite{Koning2003} to ensure that we could remove any difficulties coming from discrepancies in experimental data (such as the error in the normalization, the incident energies not lining up, etc.).  In Fig. \ref{fig:energiesMock}, we show the single and multiple optimization for $^{40}$Ca(n,n) at 13.9 MeV and $^{40}$Ca(p,p) at 14.5 MeV using mock data: confidence intervals in panels (a) and (c) and the percentage uncertainty in (b) and (d).  Comparing this figure with Fig. \ref{fig:energies} - same calculation but with real experimental data - we see that the uncertainties are smaller when mock data are used compared to the case when real experimental data are used.  In both cases though, we do not see the same degree of reduction in the uncertainty when a second energy is included in the optimization process.

%\IR{Manuel, check the proton mock data; it looks odd, especially at 12.5 MeV - the shape is the same as the real data, flat towards 0$^\circ$.}
%\IR{This has been checked and no issues found - Manuel }
%\IB{Manuel please update the legend of figure 6 (a) and (C), replace "Sim" by "mock" (for the data). Sim can get confused with Simultaneous and we always use the term mock data in the text!}

However, in \cite{CatacoraRios2019}, we also found that the uncertainty was strongly dependent on the percent difference between the two energies that are included in the fit.  There was a greater impact at higher incident energies and when the two energies included were $\sim$ 10\% apart (if the difference in energies were higher the two minima would be too different as the optical potentials are strongly energy dependent, if it were lower it would offer no extra information to the optimization procedure).  With real experimental data we cannot control this. Indeed, the existing data for this case is slightly farther apart than 10\%, which may also contribute to the differences seen here compared to \cite{CatacoraRios2019}.  Having a targeted experimental study to measure elastic scattering at two close energies could help us understand if these results are due to using real experimental data (not mock data) or from the percent difference between the two incident energies.

\section{Solving the few-body scattering problem}
\label{sec:fewbody}

%\begin{itemize}
%\item Have compared DWBA and ADWA within the same UQ framework
%\item Perform a more systematic comparison using the Bayes Factor - perhaps we can do this approximately - taking a very small error so that the uncertainties in transfer become small and doing analytic approximations in the integrand...
%\item There is probably also a Bayesian way to study the convergence of CDCC (etc.) for example, much in the same way as is done for EFT expansions - in EFT you have additional parameters with each order. in CDCC all parameters are there in first term already (n+c interaction). but i will think some more on this... \IR{I was thinking more about the uncertainties that come from convergence, not necessarily because you add more parameters.}
%\item UQ with Faddeev - if we wanted to do this we would need emulators. how would we go about developing emulators for this problem...
%\end{itemize}

\begin{figure}
\centering
\begin{tabular}{cc}
\includegraphics[width=0.5\textwidth]{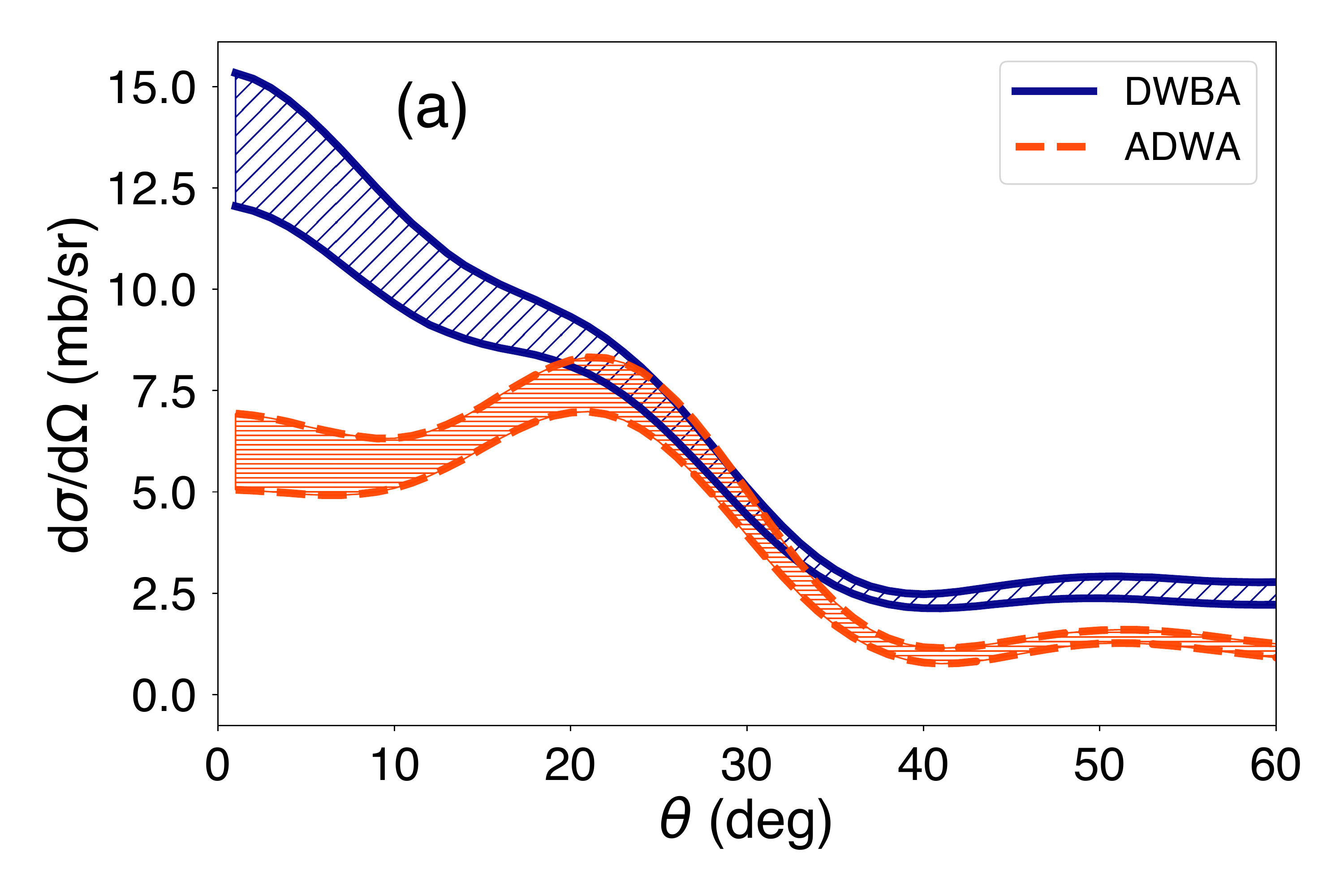} & \includegraphics[width=0.5\textwidth]{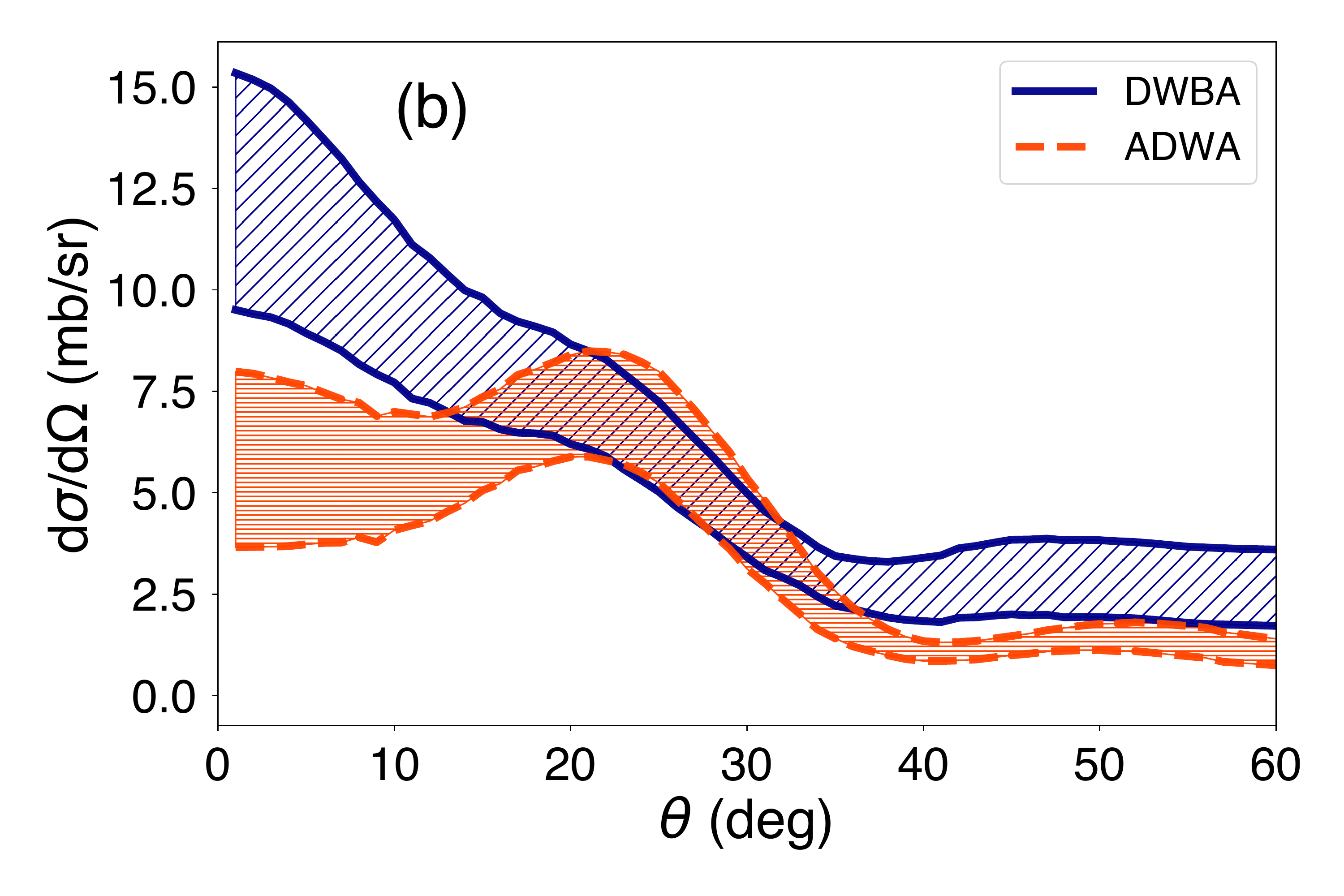}
\end{tabular}
\caption{Comparison between the $^{40}$Ca(d,p)$^{41}$Ca(g.s.) at 28.4 MeV cross section using DWBA (blue) and ADWA (red) for (a) the uncorrelated frequentist optimization and (b) the Bayesian optimization.}
\label{fig:transfer}
\end{figure}

Using the potential parameters that led to the cross sections shown in Fig. \ref{fig:UCCB}, we can propagate the uncertainties to the single-neutron transfer cross section, $^{40}$Ca(d,p)$^{41}$Ca(g.s.) at 28.4 MeV.  In this section, we consider two different three-body approximations to $(d,p)$ reaction, as discussed in Section \ref{sec:stats}, namely the distorted-wave Born approximation (DWBA) and the adiabatic wave approximation (ADWA).  In DWBA, the incoming channel requires the deuteron-target distorted wave while ADWA requires as input the proton-target and neutron-target distorted waves.  We compare DWBA and ADWA using both the uncorrelated frequentist and Bayesian optimizations, Fig. \ref{fig:transfer} (a) and (b) respectively.  Typically, a spectroscopic factor would be extracted by normalizing the calculation to the data at forward angles.  However, because there is no data near this incident energy to compare to, we consider the spectroscopic factors in each case to be unity and focus on the differences in the shape of the resulting cross sections.  

When the optical potential uncertainties are propagated through each reaction formulation, we see a significant difference in shape of the angular distributions between the DWBA and ADWA calculations, for both the frequentist and Bayesian techniques.  However, depending on the angular range over which the experimental data would be measured, these calculations would still be difficult to distinguish, unless a detailed angular distribution could be measured between zero and 40 degrees.  For stable targets, this is certainly feasible, but when considering reactions with unstable beams where the experimental errors are typically larger and the angular coverage is reduced due to the reaction being measured in inverse kinematics, these two models will become increasingly difficult to distinguish.  (Comparisons between the uncorrelated and correlated frequentist optical potentials show similar angular dependence for the DWBA and ADWA calculations, but the confidence intervals are significantly wider when the correlated fit is propagated, as would be expected from the confidence intervals on the elastic scattering in Fig. \ref{fig:UCCB}).  It is clear that the larger uncertainties on the elastic scattering calculations translate to larger uncertainties on the transfer cross section, even though these uncertainties are propagated in a non-linear fashion.  

\section{The overlap function and interplay of other degrees of freedom}
\label{sec:structure}
%AL and NC %(for preliminary results?)

%\begin{itemize}
%\item discuss ideas/things we're looking to do for these two sections
%\item Extract WS parameters for the single-particle states by fitting transfer cross sections or ANCs
%\item Amy has done a few calculations to try to get the WS parameters from a fit to (d,p) data with DWBA
%\item Have preliminary results from Nick??
%\end{itemize}

The last two items to consider are the overlap functions and the interplay of degrees of freedom that are left out of the model space.  For single-nucleon transfer reactions, when we discuss the overlap function, we are thinking in particular of the bound-state wave function between the target and the transferred nucleon.  Typically, this interaction is assumed to have a Woods-Saxon shape with a standard geometry and a depth that reproduces the nucleon-target binding energy.  In \cite{Lovell2015}, we showed that changing the geometry while keeping the binding energy fixed could drastically change the resulting transfer cross section, by a factor of two or more at the peak of the angular distribution.  These differences were seen when the radius parameter was changed from 1.1 fm to 1.3 fm.  Between the two cases that were studied, the changes were larger in the more exotic target, $^{132}$Sn, compared to the stable target, $^{48}$Ca.  

We would like to study the uncertainties in the description of the bound state in a Bayesian manner.  One option to incorporate these uncertainties would be to sample the geometry of the single-particle potential, constrained by narrow prior distributions for the radius and diffuseness, and imposing that the potential depth reproduces the binding energy.  But one could also use constraint of other known properties. The asymptotic normalization coefficient (ANC) provides one such constraint.  
Previous reaction studies have shown that breakup cross sections are scaled by the ANC squared \cite{capel2006,capel2007}.
Since the ANC can be calculated microscopically for light nuclei \cite{brida2011} and can be extracted unambiguously from peripheral reactions for many other nuclei \cite{combined}, the ANC constrain could strongly reduce the ambiguity in the singe-particle potential.
The effects of using the ANC as a constraint can be quantified with the current tools (similarly to what was done in \cite{CatacoraRios2019}). Were this to provide a significant constraint on the single-particle potential, it would give a strong motivation to develop a program to measure this quantity for a variety of heavy nuclei.  In addition, the overlap function can be extracted from the one-body density matrix (OBDM) \cite{Ivanov2001,Timofeyuk2020,Dimitrova1997}, which has already been calculated for the $^{40}$Ca target studied in this work.  Extracting the uncertainties on the overlap function using both of these methods could provide an interesting comparison between the two constraints.

The last source of uncertainty in few-body reactions discussed in Section \ref{sec:mot} comes from the degrees of freedom left out of the model space.  The most obvious source of this uncertainty is the reduction of the many-body model space to a few-body problem.  If one could compare exact many-body calculations directly with the few-body calculations, it would allow us to construct an emulator to connect the few-body  to the many-body calculations (which typically contains an uncertainty estimate such as from a Gaussian Process \cite{Rasmussen}). Unfortunately, this is currently not possible for any but the lightest systems and given the computational scale of \emph{ab-initio} reaction calculations it would take an unfeasibly long time.

Instead, what can be done effectively is to test different levels of approximations in the few-body models. We have begun to explore the differences between DWBA and ADWA calculations,  two of the few-body approximations that are commonly made.  The study discussed in Sec. \ref{sec:fewbody} explores the differences in the uncertainty interval of the predicted cross sections when one used deuteron optical potential uncertainties versus nucleon optical potential uncertainties.  A quantitative comparison of ADWA and DWBA can be performed more systematically using Bayesian evidence, as will be discussed briefly in the conclusions.

In the framework of model comparison, ADWA and DWBA are not relatable in the sense that they are not nested: one is not a simplification of the other. It is best to first consider nested models.
In that context, the coupled-channel Born approximation (CCBA) is a useful model, because one can easily switch of the coupling to the inelastic state and reduce the model to DWBA. With the Bayesian machinery that we have put in place, we can systematically study the effects that the addition of inelastic scattering channels has on the widths and means of the confidence intervals for the cross sections.  This starting point would enable the development of the methodology needed for more complex comparisons where one model is not just a subset of another.

%\IB{Fig. 5, constraining the single-particle state parameters either from (d,p) or ANC}

%\section{Interplay of other degrees of freedom}
%\label{sec:dof}
%FN and MC
%\begin{itemize}
%\item Use single channel OM for elastic and then couple channel OM elastic/inelastic
%\item Use Bayes Factor to assess the need for introducing excited states
%\end{itemize}

%\IB{Fig. 6, elastic/inelastic with various channels included}

\section{Outlook}
\label{sec:outlook}

%\begin{itemize}
%\item summary of the current work
%\item describe a vision for experimental design
%\item model comparison
%\item model mixing
%\item using Bayesian evidence
%\end{itemize}

In this overview, we have described, using the example of $^{40}$Ca elastic scattering and transfer reactions, recent progress made on quantifying uncertainties in few-body reaction theory.  We have shown the development from frequentist $\chi^2$ minimization methods and covariance matrix propagation to a full Bayesian analysis in determining and propagating uncertainties in effective interactions.  In addition, under these two frameworks, we can now directly compare how uncertainties are propagated from these interactions when various approximations to the few-body model are made, namely between the distorted-wave Born approximation and the adiabatic wave approximation.  Although much work has been done, there are many exciting opportunities ahead.  The long-term vision for the future of this work involves using Bayesian methodology along two major interacting thrusts: the first concerns experimental design and second is focused on improving theory itself (one might call it theory design).  In this section we discuss the overarching vision and specific developments that are needed for implementing that vision.

One tool that we have identified as necessary to implement this long-term vision is calculating the Bayesian evidence for various theories.  Described briefly in Sec. \ref{sec:theory}, the Bayesian evidence is the denominator in Bayes' theorem, Eq. (\ref{eqn:bayes}), which is a weighted sum over all possible models and provides a formal way of evaluating relative probabilities of different models.  Bayesian evidence is a numeric formulation of \emph{Occam's razor} - a systematic way of showing that a simpler model which reproduces the data should be favored over a more complex model.  However, computing the Bayesian evidence typically requires an integral over all of a model's parameter space, a potentially computationally demanding task, particularly as the models become more complex.  In addition, the interpretation of the value of the evidence is not always straight-forward.  Still, being able to calculate the Bayesian evidence is an important development for our long-term uncertainty quantification vision.

In the first thrust of experimental design, there are many controllable conditions in reaction experiments (beam energy, angular range, target, etc.), and the tools developed thus far can help in determining those conditions that produce optimum sensitivity to the quantities of interest, e.g. as in \cite{Melendez2020}. The work in \cite{CatacoraRios2019}  is the first step along this direction, but other tools should be considered when assessing what combination of observables contains most information. Tools such as Principle Component Analysis (PCA) and Bayesian evidence have been successfully applied in other areas, e.g. \cite{Sangaline2016,Novak2014,Trotta2008,Fox2020}, and should be studied in the context of nuclear reactions.  

The Bayesian methodology here discussed can also be extremely useful in improving theoretical models.
So far, due to computational considerations, the models used have been very simple (optical model, DWBA, and ADWA). However, we understand that there is important physics that these models do not contain. Ultimately, we want to increase the complexity of the models, matching the state of the art in theory to the degree called for by the data. In other words, the model should be as sophisticated at the data requires. In addition, as we include more and more data in our analysis, there will be an increasing demand on the physics included in the model to be able to describe all observables simultaneously with the same accuracy. In order to start down this path, we will need quantitative and reliable methods for model comparison. An essential tool for performing model comparisons in the Bayesian world is the Bayesian evidence. Also, model mixing is the natural sequence from model comparison, as a robust way to incorporate the best aspects of each theory.  Both of these developments, though, pose specific challenges that still need to be addressed. What to do when the models are not nested, and moreover when the models do not contain similar parameters?  What are the most efficient and accurate methods to perform numerical integrals over a large parameter space?  These are questions that must be investigated to ensure a robust development in this area.

%\IR{Still need a chart/figure}

%\vspace{1cm}
%\small
\ack
This work was supported by the National Science Foundation under Grant PHY-1403906 and the Department of Energy under Contract No. DE-FG52-08NA28552 and was performed under the auspice of the U.S. Department of Energy by Los Alamos National Laboratory under Contract 89233218CNA000001.  A.E.L. would like to thank W.J. Ong for her help in brainstorming names for the {\sc QUILTR} wrapper.  We also gratefully acknowledge the support of the U.S. Department of Energy through the LANL/LDRD Program and the Center for Non Linear Studies.  

\newpage
\normalsize
\bibliography{jpg-expth}

\end{document}